  \documentclass{amsart}
  \usepackage{url}
  \usepackage{amssymb}
  \input xypic
  \xyoption{all}

  \usepackage{amsmath, amsthm,  amsfonts}

  \theoremstyle{plain}
  \newtheorem{theorem}{Theorem}[section]
  \newtheorem{lemma}[theorem]{Lemma}
  \newtheorem{proposition}[theorem]{Proposition}
  
  \newtheorem{corollary}{Corollary}[section]

  \theoremstyle{definition}
  \newtheorem{definition}[theorem]{Definition}

  \theoremstyle{remark}
  \newtheorem{remark}[theorem]{Remark}
  \newtheorem{example}{Example}[section]
  \newtheorem{note}{Note}[section]

  \numberwithin{equation}{section}
  \numberwithin{figure}{section}

  \renewcommand{\cH}{{\mathcal H}}

  \newcommand{\cQ}{{\mathcal Q}}
  \newcommand{\cA}{{\mathcal A}}
  \newcommand{\cE}{{\mathcal E}}
  \newcommand{\cM}{{\mathcal M}}
  \renewcommand{\cL}{{\mathcal L}}
  \renewcommand{\cD}{{\mathcal D}}
  \newcommand{\cP}{{\mathcal P}}
  \newcommand{\cU}{{\mathcal U}}
  \newcommand{\cG}{{\mathcal G }}
  \newcommand{\cC}{{\mathcal C }}
\newcommand{\cO}{{\mathcal O }}
\newcommand{\cZ}{{\mathcal Z }}
\newcommand{\cN}{{\mathcal N }}

   \newcommand{\ba}{\begin{eqnarray}}
   \newcommand{\na}{\end{eqnarray}}
   \newcommand{\ban}{\begin{eqnarray*}}
   \newcommand{\nan}{\end{eqnarray*}}

  \newcommand{\g}{{\mathfrak g}}
  \renewcommand{\t}{{\mathfrak t}}

  \newcommand{\CC}{{\mathbb C}}
  \newcommand{\RR}{{\mathbb R}}
  \newcommand{\ZZ}{{\mathbb Z}}
  
  \renewcommand{\AA}{{\mathbb A}}

  \newcommand{\PP}{{\mathbb P}}

  \renewcommand{\a}{\alpha}

    \newcommand{\disp}{\displaystyle}

  \def\cancel#1#2{\ooalign{$\hfil#1\mkern1mu/\hfil$\crcr$#1#2$}}

\def\dirac{\mathpalette\cancel\partial}

\begin{document}

  \title[ Fusion   of  symmetric $D$-branes and Verlinde rings]
{ Fusion   of  symmetric $D$-branes and Verlinde rings }

  \author[A.L. Carey]{Alan L. Carey}
\address[Alan L. Carey]
  {Mathematical Sciences Institute\\
  Australian National University\\
  Canberra ACT 0200 \\
  Australia}
  \email{acarey@maths.anu.edu.au}

  \author[Bai-Ling Wang]{Bai-Ling Wang}
  \address[Bai-Ling Wang]
  {Mathematical Sciences Institute\\
  Australian National University\\
  Canberra ACT 0200 \\
  Australia}
  \email{wangb@maths.anu.edu.au}

  \thanks{The authors acknowledge the support of the Australian
  Research Council. BLW  thanks Valerio Toledano Laredo for stimulating
  e-mail correspondence. ALC thanks
the Erwin Schrodinger Institute for their  support.}

  \begin{abstract} We explain how
  multiplicative bundle gerbes over a compact, connected  and
   simple Lie group $G$ lead to a certain fusion category
   of equivariant bundle gerbe modules given by pre-quantizable Hamiltonian
   $LG$-manifolds arising from Alekseev-Malkin-Meinrenken's
   quasi-Hamiltonian $G$-spaces. The motivation comes from string theory
namely,
by generalising the notion of $D$-branes in $G$
to allow subsets of $G$ that are the image of
a $G$-valued moment map we can define a `fusion of $D$-branes' and a map
to the  Verlinde ring of the loop group of $G$ which preserves the product
structure.   The idea is suggested by the theorem of Freed-Hopkins-Teleman.
The case where $G$ is not simply connected is studied carefully in terms of
equivariant bundle gerbe  modules for multiplicative bundle gerbes.
  \end{abstract}
  \maketitle

\tableofcontents

  \section{Introduction}

 It was shown in \cite{BCMMS} that there is
an additive group structure on equivalence classes
of bundle gerbe modules, for
a bundle gerbe over a manifold $M$ with Dixmier-Douady class
$[H]\in H^3(M, \ZZ)$, such that the resulting group is isomorphic
to the twisted K-theory of $M$ twisted by $[H]$, denoted $K_{[H]}(M)$.
 On the other hand the
 theorem in \cite{FHT} is that the Verlinde ring
of positive energy representations of the loop group
of a compact Lie group (with fusion $*$ as the product
and denoted $(R_k(LG), \ast)$) is isomorphic to
 the equivariant twisted K-theory $K^{dim G}_{G, k+ h^\vee}(G)$ where $h^\vee$
 is the dual Coxeter number. Here $k+ h^\vee$ is viewed as the level of the twisting class in
 $H^3_G(G)$ with the $G$-action on $G$ given  by conjugation.

The aim of this paper is to answer
the natural question: is there is a fusion product which can
be constructed using
bundle gerbe modules and a direct  map to  $(R_k(LG), \ast)$
preserving the fusion product structure.  For simplicity,
we only deal with a bundle
gerbe over a compact, connected and simply-connected
   simple Lie group in this introduction.
   We have some results on the general situation in Section 6.

Our approach depends on a second circle of ideas.
First, $(R_k(LG), \ast)$ provides the quantization of
classical Wess-Zumino-Witten models with target $G$. Second,
bundle gerbes over $G$ provide a differential geometric way
to approach Wess-Zumino-Witten models \cite{CMM}. Third,
the main result of \cite{CJMSW} shows that
classical Wess-Zumino-Witten models which arise
by transgression from Chern-Simons
gauge theories have the property that their
 associated bundle gerbe has internal extra structure
termed  `multiplicative'.
In fact we showed more namely that a bundle gerbe $\cG$ over $G$ is
multiplicative (see Theorem 5.8  in \cite{CJMSW}),
  if and only if its Dixmier-Douady class is transgressive, that is,
lies in the image of
the transgression map $\tau: H^4(BG, \ZZ) \longrightarrow H^3(G, \ZZ).$
We will see that the multiplicative property gives a
fusion product for bundle gerbe modules.
For a simply-connected, compact simple Lie group $G$, we know
that $H^4(BG, \ZZ)\cong H^3(G, \ZZ) \cong \ZZ$
and so for any integer $k \in H^3(G, \ZZ) \cong \ZZ$
there is a corresponding multiplicative bundle gerbe $\cG_k$.
To assist the reader we review the key notions from \cite{CJMSW}
in subsection 4.1.

While it is not necessary in order
to understand the mathematical results of this
paper, our motivation comes from string theory considerations. Namely
we expand the notion of a $D$-brane (as there is no fusion product on
the space of $D$-branes) to include
 the image of $G$-equivariant smooth maps
from a manifold to $G$ (Cf. Definition \ref{bg:$D$-brane}). We are able
using the multiplicative property, to define fusion for these
generalised $D$-branes.

\subsection{Some physics background}

This subsection is not really needed for our results but
we use it to introduce some notation and
for the interested reader we include some background which amplifies
the preceding remarks. First we recall that the
  background Kalb-Ramond field (the so called ``B-field'')
   is a 2-form potential for an invariant 3-form
  on the target group manifold $G$.
In Type II string theory with non-trivial \lq\lq$B$-field", twisted
 K-theory is believed to classify those $D$-branes with Chan-Paton fields
(\cite{Wit1}).

    The quantized Wess-Zumino-Witten
  model of level $k$ (a positive integer) for  closed strings moving on a
  group manifold $G$ is  determined by
  a closed string Hilbert space
  \[
  \cH_{WZW}^k = \sum_{\lambda\in \Lambda_k^*} H_{*\lambda}\otimes H_\lambda,
  \]
   where $H_\lambda$ is the positive energy irreducible  projective representation
   of the loop group $LG$ at level $k$ with dominant weight $\lambda$ in the space of level
   $k$ dominant weights $\Lambda_k^*$
 and $*\lambda$ is the dominant weight of the irreducible representation of $G$
   complex conjugate to the one with weight $\lambda$. In addition there is
an assignment of
    trace class operators
  \[
  Z_k(\Sigma): \underbrace{\cH_{WZW}^k \otimes \cdots \otimes\cH_{WZW}^k}_{\text{$m$ times}}
 \longrightarrow \underbrace{\cH_{WZW}^k \otimes \cdots \otimes\cH_{WZW}^k}_{\text{$n$ times}}
\]
 to any Riemann surface  $\Sigma$ with analytically
  parametrised boundaries divided into
   $m$ incoming boundaries and
  $n$ outgoing boundaries, such that the operators $Z_k(\Sigma)$ satisfy
  certain gluing formulae under composition of surfaces (See \cite{Seg} and the
   references therein).
  We call such Riemann surfaces `extended'.

   With primary fields, determined by a dominant weight
  of level $k$, inserted at each boundary of $\Sigma$
the space of correlations or conformal
  blocks is given by the multiplicity space of
  the modular functor from the category of extended Riemann
  surfaces with conformal structure to
  the category of positive energy irreducible  projective representations
   of the loop group $LG$ at level $k$:
   \[
   \cH_\Sigma = \sum_{\overrightarrow{\lambda}_{in}, \overrightarrow{\lambda}_{out}}
   V_{\Sigma}^k(*\overrightarrow{\lambda}_{in}, \overrightarrow{\lambda}_{out})
    \otimes  H_{*\overrightarrow{\lambda}_{in}}
   \otimes H_{ \overrightarrow{\lambda}_{out}}.
   \]
   The space of conformal blocks
   $V_{\Sigma}^k(*\overrightarrow{\lambda}_{in}, \overrightarrow{\lambda}_{out})$
   also satisfies certain gluing formulae, the well-known Verlinde
   factorization formulae. Varying the conformal structure on $\Sigma$, the
   space of conformal blocks forms a holomorphic vector bundle over the
   moduli space of conformal structure, equipped with a canonical projective
   flat connection (the Knizhinik-Zamolodchikove connection).

   For $\Sigma_{0, 3}$,  the genus $0$ surface with 3   boundary components,
two incoming
   boundary circles labelled by the weight $\lambda$, $\mu$ and the third
outgoing
   boundary circle labelled by the weight $\nu$, the dimension of the space
   of conformal blocks
   \ba\label{Verlinde:coeff}
   N_{\lambda,\mu}^{\nu}
   = dim V_{\Sigma_{0, 3}}^k(*\lambda, *\mu, \nu)
   \na
    is given by the Verlinde fusion coefficient (\cite{Ver}). Another
     definition of Verlinde coefficients $N_{\lambda\mu}^\nu$
     is given by (Cf. \cite{Bea}):
\[
N_{\lambda,\mu}^\nu =dim \{ u\in Hom_{G}(V_\lambda \otimes
V_{\mu}, V_\nu)| u\bigl(V_\lambda^{(p)} \otimes
V_{\mu}^{(q)}\bigr) \subseteq \bigoplus_{p+q+r \leq k}
V_{\nu}^{(r)}\}
\]
where $V_\lambda, V_{\mu}$ and $ V_\nu $ denote the representation
of $ G$ with highest weight $\lambda$, $\mu$ and $\nu$
respectively. The highest root $\vartheta$ determines a copy of
$SU(2)$ with respect to which $V_\lambda$  admits a decomposition
\[
V_\lambda =\sum_{i=0}^{k/2} V_\lambda^{(i)}
\]
where $V_\lambda^{(i)}$ ($i=0, 1/2, 1, \cdots, k/2$) are the spin
$i$ isotypic components.

    In boundary conformal field theory, the $D$-brane is described by
    a boundary state in the closed string Hilbert space, which
    is a linear combination of the so called Ishibashi states.
    The coefficients should satisfy the Cardy condition and some
    sewing relations (Cf. \cite{Car}).
For  the  Wess-Zumino-Witten model of conformal field theory,
 the stringy geometry can be studied via
  bundle gerbes (\cite{Gaw}\cite{GawRei})
and embedded
submanifolds $Q$ of $G$, the $D$-branes.
  For the boundary Wess-Zumino-Witten theory on a simply connected
  group manifold $G$ (Cf. \cite{GawRei}),
  symmetry preserving boundary
  conditions are labelled by $\lambda\in \Lambda^*_k$ and the open
  string Hilbert space labelled by $\lambda_1$ and $\lambda_2$
  admits the following decomposition
  \[
  \cH_{\lambda_1, \lambda_2}^{open} \cong \sum_{\mu \in \Lambda^*_k}
  W_{\lambda_1\mu}^{\lambda_2}\otimes H_\mu.
  \]
  In order to get a consistent quantum conformal field  theory,
   the multiplicity space
  $W_{\lambda_1\mu}^{\lambda_2}$ is identified with the
  space of conformal blocks $V_{\Sigma_{0, 3}}^k(*\lambda_1, *\mu, \lambda_2)$. In particular,
  the dimension of the multiplicity space
  \[
  dim  \ W_{\lambda_1\mu}^{\lambda_2} = N_{\lambda_1,\mu}^{\lambda_2}.
  \]
  is also   given by the Verlinde fusion coefficients. For  general boundary conditions,  the consistency
  condition implies that 
  $\mu\mapsto (dim  \ W_{\lambda_1\mu}^{\lambda_2})$ realises a representation of the Verlinde
  algebra.
See also \cite{AleSch} \cite{BDR} \cite{FFFS} \cite{FS}  \cite{GGR} for some earlier
discussion of $D$-branes on group manifolds. Geometrically, for
a simply-connected Lie group $G$,  $D$-branes on $G$ are classified into
symmetric $D$-branes:
\[
\cC_\lambda =\{ g\cdot exp (\disp{\frac{2\pi i\lambda}{k}})) \cdot g^{-1} | g\in G\}
\]
for $\lambda \in \Lambda_k^*$;  twisted $D$-branes
\[
\cC_\lambda =\{ g\cdot exp (\disp{\frac{2\pi i\lambda}{k}})) \cdot \varpi_G(g)^{-1} | g\in G\}
\]
where $\varpi_G$ is an outer automorphism of $G$, and $\lambda\in \Lambda_k^*$
is a fixed point of $\varpi_G$, and the coset $D$-branes such as $D$-brane s in 
$\cN=2$ coset models in \cite{LW}\cite{MMS}\cite{Sch1}\cite{Sch2}.  
In this paper, we will study these symmetric $D$-branes
from the equivariant bundle gerbe module viewpoint.

\subsection{Mathematical summary}

We now look at the mathematical content of this paper.
We introduce a new notion,
`generalized rank $n$ bundle gerbe $D$-branes' for a bundle
gerbe $\cG$ over $G$. These are smooth manifolds $Q$ with a smooth
map $\mu:Q\to G$ such
that the pull-back bundle gerbe $\mu^*(\cG)$
admits a rank $n$ bundle gerbe module
(Definition \ref{bg:$D$-brane}). There is also
a corresponding notion for $G$-equivariant bundle
gerbes $\cG$ over a $G$-manifold $M$.

 When the compact simple Lie group $G$  is simply-connected
then we can construct a
$G$-equivariant bundle gerbe $\cG_k$
over $G$ whose Dixmier-Douady class is represented by
a multiple by a positive integer $k$ of the
canonical bi-invariant 3-form on $G$
(see  Proposition \ref{lift:G-equivariant} in Section 3).

A particularly interesting example of a generalized $G$-equivariant  bundle
gerbe $D$-brane is provided
by a quasi-Hamiltonian manifold $(M, \omega, \mu)$
(see Definition \ref{quasi-hamiltonian}) where $M$ is a $G$-manifold,
$\omega$ is
an invariant 2-form and $\mu:M\to G$ is a group-valued moment map.
 Quasi-hamiltonian manifolds are extensively
studied by Alekseev-Malkin-Meinrenken in \cite{AAM} whose results are reviewed
in  Section 3.
We focus on the correspondence
 between  quasi-Hamiltonian manifolds and Hamiltonian $LG$-manifolds at level $k$
as illustrated by the following diagram:
 \ba\label{G:LG}
 \xymatrix{ \hat{M}\ar[d]_{\pi} \ar[r]^{\hat{\mu}} & L\g^* \ar[d]^{Hol}\\
     M \ar[r]^{\mu} &G,
     }
     \na
  where $\hat{\mu }:   \hat{M} \to L\g^*$ is the moment map for the
   Hamiltonian $LG$-action at level $k$ and the vertical arrows
 define $\Omega G$-principal bundles.  The  quasi-Hamiltonian manifold,
when ``pre-quantizable'', is naturally a generalized
rank 1 bundle gerbe $D$-brane (Cf. Theorem \ref{bg:module:rank1}) of the
bundle gerbe over $G$.

When $G$ is semisimple and
simply connected, any   bundle gerbe $\cG_k$ is multiplicative \cite{CJMSW}
and so in Section 4, applying
Theorem \ref{bg:module:rank1} to the   moduli spaces
of flat connections on Riemann surfaces, we relate these generalized
bundle gerbe $D$-branes
to the Chern-Simons bundle 2-gerbe of  \cite{CJMSW} over the classifying space
 $BG$ .
A corollary is that the Segal-Witten reciprocity law is explained in the set-up
 of multiplicative bundle gerbes and their generalized bundle gerbe $D$-branes.

In Section 5, we define the fusion
category of generalized bundle gerbe $D$-branes of $\cG_k$
to be the category of pre-quantizable quasi-Hamiltonian manifolds with fusion
product
\[
(M_1, \omega_1, \mu_1)\boxtimes (M_2, \omega_2, \mu_2) =
(M_1 \times M_2, \omega_1 + \omega_2 + \disp{\frac k2} <\mu^*_1\theta, \mu_2^*\bar{\theta}>,
 \mu_1\cdot \mu_2),
\]
where the $G$-action on $M_1\times M_2$ is via the diagonal embedding
$G \to G\times G$, $\theta$,  $\bar{\theta}$ are the left and right Maurer-Cartan forms
on $G$, and $\mu_1\cdot \mu_2(x_1, x_2) = \mu_1(x_1)\cdot \mu_2(x_2)$.
This fusion product and the corresponding fusion product
on Hamiltonian $LG$-manifolds were studied in \cite{MeiWoo1}.
Denote by $(\cQ_{G, k}, \boxtimes)$ the fusion category of bundle gerbe
 $D$-branes of $\cG_k$.

  Let $R_k(LG)$ be the free group  over $\ZZ$ generated by  the isomorphism classes of
  positive energy,    irreducible, projective representations of $LG$
     at level $k$.
The central extension of $LG$ at level $k$ we write as $\widehat{LG}$.
The positive energy representation labelled
by $\lambda\in \Lambda_k^*$ acts on  $\cH_\lambda$ and
    the Kac-Peterson character of $\cH_\lambda$  is
    \ba\label{kac:character}
        \chi_{k, \lambda} (\tau) = Tr_{\cH_\lambda} e^{2\pi i \tau (L_0 -\frac{c}{24})},
    \na
    where $\tau \in \CC$ with $Im (\tau) >0$,
 $L_0$ is the energy operator on
    $\cH_\lambda$ (Cf.\cite{PS}), and $c= \disp{\frac{kdim G} {k + h^\vee}}$ is the
    Virasoro central charge.  We mention that
     $e^{2\pi i \tau (L_0 -\frac{c}{24} )}$ is a trace class operator (Cf. Theorem 6.1
     in \cite{GW} and Lemma 2.3 in \cite{EFK}) for  $\tau \in \CC$ with $Im (\tau) >0$.
      Equipped with  the fusion ring structure:
\[
\chi_{\lambda,k} \ast \chi_{\mu, k} = \sum_{\nu\in
\Lambda_k^*}N_{\lambda,\mu}^{\nu}\chi_{\nu, k},
\]
where $N_{\lambda,\mu}^{\nu}$ is the Verlinde fusion coefficient
(\ref{Verlinde:coeff}), we obtain $(R_k(LG), \ast)$,  the Verlinde
ring.

 Motivated by Guillemin-Sternberg's ``quantization commutes with reduction"
philosophy, we define a quantization
 functor on the fusion category of generalized bundle gerbe $D$-branes of $\cG_k$
 using $Spin^c$ quantization of the reduced spaces (see Definition \ref{quantization}):
 \[
 \chi_{k, G}: \cQ_{G, k} \longrightarrow R_k(LG).
 \]
 Note that for a quasi-Hamiltonian manifold $M$ obtained from
 a pre-quantizable Hamiltonian $G$-manifold, $\chi_{k, G}(M)$ is
 the equivariant index of the $Spin^c$ Dirac operator twisted
 by the pre-quantization line bundle.

 \noindent    {\sl {\bf  Main Theorem}: The quantization  functor
$\chi_{k, G}: (\cQ_{G, k}, \boxtimes)  \longrightarrow (R_k(LG), \ast)$
satisfies
\[
\chi_{k, G} ( M_1\boxtimes M_2) = \chi_{k, G} (M_1 )\ast\chi_{k, G} (M_2 ),
\]
where the product $\ast$ on the right hand side denotes the fusion ring
structure
on the Verlinde ring $(R_k(G),*)$.}

The fusion product
on Hamiltonian $LG$-manifolds  at level $k$ involves the moduli space of flat connections on
 a canonical pre-quantization line bundle over the `trousers' $\Sigma_{0, 3}$.
The multiplicative property of the bundle gerbe $\cG_k$
over $G$ is essential for this part of the construction.

  In section 6, we   discuss various subtle
  issues concerning the non-simply connected case.
Given a compact, connected,
non-simply connected simple Lie group $G=\tilde G/Z$  for a
subgroup $Z$ in the center $Z(\tilde G)$ of
the universal cover $\tilde G$, we construct
a  $G$-equivariant bundle gerbe $\cG_{(k, \chi), G}$  associated to
a multiplicative level $k$ and a character $\chi \in Hom(Z, U(1))$,
where   the so-called level  lies in $H^4(B\tilde G, \ZZ)$ and $k$ is the
multiplicative if it is transgressed from $H^4(BG, \ZZ)$ to $ H^4(B\tilde G, \ZZ)$.

The $G$-equivariant bundle gerbe $\cG_{(k, \chi), G}$ is
obtained from the central  extension of $LG$ in \cite{Tol},
$1\to U(1) \longrightarrow \widehat{LG}_\chi \longrightarrow LG \to 1,$
associated to $(k, \chi)$. We   classify all
irreducible positive energy representations of
$\widehat{LG}_\chi$ following the work of  Toledano Laredo in \cite{Tol}.

Let $R_{k, \chi}(LG)$ be the
Abelian group generated by the positive energy, irreducible representations
of $\widehat{LG}_\chi$.  We define  the category
$\cQ_{(k, \chi), G}$ of $G$-equivariant bundle gerbe
modules of $\cG_{(k, \chi).  G}$.
The quantization functor
\[
\chi_{(k, \chi), G}: \cQ_{(k, \chi), G} \longrightarrow
R_{k, \chi}(LG)\]
can also be established (See Definition \ref{quantization:non}).

When $k$ is multiplicative and $\chi$ is the trivial homomorphism $1$, then
 $\cQ_{(k, 1), G}$ admits a natural fusion
product structure whose resulting category is denoted by
$(\cQ_{(k, 1), G}, \boxtimes )$. Then $ \chi_{(k,
1), G}$ induces a ring structure on $R_{k, 1}(LG)$ (Cf. Theorem \ref{main:non}).

   \section{Bundle gerbe $D$-branes}

  It is now known that the ``B-fields" on a manifold $M$ can be described by a bundle
  gerbe with connection and  curving, and topologically classified
  by the degree 2 Deligne cohomology $H^2(M, \cD^2)$. (This was understood
in \cite{BCMMS} using
\cite{FreWit}). Explicitly,
choose a good covering $\{U_i\}$ of $M$. Denote double intersections
$U_i\cap U_j$ by $U_{ij}$ and extend this notation  in the obvious way to
$n$-intersections. Then a degree 2 Deligne cohomology
  class is given by an equivalence class of triples
  \[
  (g_{ijk}, A_{ij}, B_{i})
  \]
  where $g_{ijk}\in C^\infty(U_{ijk}, U(1))$, $A_{ij}
  \in \Omega^1(U_{ij}, i\RR)$ and $B_{i}\in \Omega^2(U_i, i\RR)$
  satisfy the following cocycle condition
  \[
  \begin{array}{lll}
  &g_{ijk}g^{-1}_{ijl}g_{ikl}g_{jkl}^{-1}   = 1, \quad & \text{ on $U_{ijkl}$} \\[2mm]
  & A_{ij}+A_{jk}+A_{ki} = g_{ijk}^{-1}dg_{ijk}, &  \text{ on $U_{ijk}$} \\[2mm]
  & B_i -B_j = d A_{ij}&  \text{ on $U_{ij}$}.
  \end{array}
  \]
  The equivalence relation is given by adding a coboundary term
  \[
  (h_{ij}h_{jk}h_{ki}, A_j-A_i - h_{ij}^{-1}d h_{ij}, 0)
  \]
  for $h_{ij} \in C^\infty(U_{ij}, U(1))$ and $A_i \in \Omega^1(U_i, i\RR)$.

  Differential geometrically, a degree 2 Deligne cohomology
  class can be realized by  a bundle gerbe with connection and curving
  over $M$ \cite{Mur}. A bundle gerbe $\cG$ over $M$ consists of a quadruple
  $(\cG, m; Y, M)$ where $Y$ is   a smooth
 manifold  with a
 surjective submersion $\pi:  Y\to M$, and  a principal $U(1)$-bundle (also denoted
 by $\cG$)
  over the fibre product
$Y^{[2]}= Y\times_\pi Y$ together with a groupoid multiplication $m$
  on $\cG$, which is compatible with the natural  groupoid multiplication
  on $Y^{[2]}$. We represent a bundle gerbe $\cG =(\cG, m; Y, M)$
  by the following diagram
  \ba
  \xymatrix{
 \cG \ar[d]& \\
  Y^{[2]}\ar@< 2pt>[r]^{\pi_1} \ar@< -2pt>[r]_{\pi_2}
  & Y   \ar[d]\\
   &M
   }
   \label{bg}
   \na
   with the bundle gerbe product $m$ given by an isomorphism
   \ba\label{bg:product}
   m:  p^*_1\cG \otimes p^*_3\cG   \to p^*_2\cG
   \na
   of principal $U(1)$-bundles over $Y^{[3]} = Y\times_\pi Y\times_\pi Y$, and
       $p_i,\ i=1,2,3$  are the three natural projections from
   $Y^{[3]}$ to $Y^{[2]}$ obtained by omitting the entry in  position $i$ for $p_i$. The maps $\pi_j$,
$j=1,2$
are the projections onto the first and second factors in $Y^{[2]}$.

   A bundle gerbe with connection and curving $\cG$ over $M$
   is given by (\ref{bg}), together with  a $U(1)$-connection $A$ on
   the principal $U(1)$-bundle $\cG$ over $Y^{[2]}$
   which is compatible  with the bundle gerbe product $m$:
   \[
   m^*(p_2^* A) = p_1^*A + p_3^*A,
   \]
   and a 2-form $B$ on $Y$, such that the curvature $F_A$ of $A$,
satisfies the relation
   $$F_A = \pi_1^*(B)-\pi^*_2(B).$$
   The connection $A$ is called a bundle gerbe connection and
     $B$ is called the curving of $A$.
     Then there exists a closed
   3-form $H$ called the bundle gerbe curvature of $(\cG, A)$,
    such that $dB = \pi^*H$. (The notation is chosen to match with the
corresponding objects in the Deligne point of view.)

The characteristic class of the Deligne class $[(g_{ijk}, A_{ij}, B_{i})]$
  is given by the class of the  \v{C}ech cocycle $\{g_{ijk}\}$, in
  \[
  H^2(M,  \underline{U(1)}) \cong H^3(M, \ZZ).
  \]
  The corresponding class in $H^3(M, \ZZ)$ for any realizing bundle gerbe
  $\cG$ is called the Dixmier-Douady class    of $\cG$.

   In \cite{BCMMS}, bundle gerbe modules are defined to study
 twisted K-theory.  Given a bundle gerbe $\cG = (\cG, m; Y, M)$ over $M$, a rank $n$  bundle
    gerbe module of $\cG$ is a   rank $n$ Hermitian vector bundle $\cE$ over $Y$,
  associated to a $U(n)$-principal bundle $P$ over $Y$ for which
    there is an isomorphism of principal  bundles  over $Y^{[2]}$
    \ba\label{bg:module}
    \rho: \cG \otimes \pi^*_2 P \cong \pi^*_1 P
    \na
    which is compatible with the bundle gerbe product:
    \[
    \rho \circ ( m \otimes id) = \rho \circ (id \otimes \rho).
    \]
    Note that a bundle gerbe $\cG$ admits a rank $n$  bundle
    gerbe module $(\cE, \rho)$ if and only if the Dixmier-Douady class
    of $\cG$ is a torsion class in $H^3(M, \ZZ)$.

    \begin{remark}
    The next definition is motivated by the following stringy considerations.
     In Type II superstring theory with non-trivial B-field  on a 10-dimensional
    oriented, spin manifold $M$, a $D$-brane $Q$,  as defined in \cite{FreWit} \cite{Wit1},
     is given by a smooth oriented submanifold
$    \iota:Q \rightarrow M$
    such that
    \[
    \iota^*z_B + W_3(Q) =0,
    \]
    where $z_B \in H^3(M, \ZZ)$ is the characteristic class of
    the B-field, and $W_3(Q)$ is the third integral Stieffel-Whitney class,
    the obstruction to the existence of a $Spin^c$ structure on $Q$.
    Now there is
     a torsion bundle gerbe $\cG_{W_3}$ (with the Dixmier-Douady class $W_3(Q)$)
called the lifting bundle
    gerbe. It arises \cite{Mur}  from the central extension
$    1\to U(1) \longrightarrow Spin^c \longrightarrow SO \to 1.$
    Denote by $\cG_B$ the bundle gerbe determined by the B-field on $M$.
Now if  $W_3(Q)=0 $ (that is, $Q$ is a $Spin^c$
    manifold), then $\iota^*z_B=0$ so that $\iota^*\cG_B$
    admits a trivialization.   \end{remark}

    \begin{definition}
    \label{bg:$D$-brane}
     Let $\cG$ be a bundle gerbe over a manifold
    $M$ equipped with a bundle gerbe connection and curving.
    \begin{enumerate}\item A {\bf rank 1 bundle gerbe $D$-brane}
     of a bundle gerbe $\cG$ over $M$ is
    a smooth oriented submanifold
     $\iota: Q \rightarrow M$
    such that $\iota^*\cG$ admits a trivialization. Given a bundle
    gerbe connection and curving on $\cG$, a twisted gauge field on the
    $D$-brane is a   trivialization of the corresponding Deligne class.
    \item A {\bf rank $n$ bundle gerbe $D$-brane}
     of a bundle gerbe $\cG$ over $M$ is
    a smooth oriented submanifold
    $\iota:  Q \rightarrow M$
    such that $\iota^*\cG$ admits a rank $n$ bundle gerbe module. A twisted gauge field on the
    $D$-brane is a bundle gerbe module connection on the bundle gerbe module.
    \item A {\bf generalized rank $n $ bundle gerbe $D$-brane} is a smooth manifold $Q$
    with a smooth map
    $    \mu: Q  \rightarrow M$
    such that $\mu^*\cG$ admits a rank $n$ bundle gerbe module.
    A twisted gauge field on the
    $D$-brane is a bundle gerbe module connection on the bundle gerbe module.
    \end{enumerate}
    \end{definition}

\subsection{Equivariant bundle gerbe $D$-branes}
Now we recall  the definition of an equivariant bundle gerbe
from \cite{Mei} (see also \cite{MatSte}).
Let $M$ be a smooth $G$-manifold, acted on by $G$ from the left.  A $G$-equivariant
bundle gerbe over a $G$-manifold $M$ is a bundle
gerbe $(\cG, m; Y, M)$ where $Y$ is   a smooth
 $G$-manifold  with a $G$-equivariant
 surjective submersion $\pi:  Y\to M$, and  a  $G$-equivariant
 principal $U(1)$-bundle (also denoted
 by $\cG$)   over the fibre product
$Y^{[2]}= Y\times_\pi Y$ together with a $G$-equivariant groupoid multiplication $m$
  on $\cG$. Note that the diagonal embedding of $G$ into $G^p$ defines an action of $G$ on
  \[
 Y^{[p]}  =  \underbrace{Y\times_{M}Y\times_{M}\cdots\times_{M}Y }_{\text{$p$ times}}.
\]
A $G$-equivariant bundle gerbe $\cG$ defines a bundle gerbe
\[
(EG\times_G \cG, m, EG\times_G Y, EG\times_G M)
\]
over $EG\times_G M$ whose Dixmier-Douady
class defines  an element in
\[
H^3_G (M, \ZZ) = H^3( EG\times_G M, \ZZ),
\]
called the equivariant Dixmier-Douady
class of $\cG$.

 Conversely, given an element of  $H^3_G (M, \ZZ)$, Section 6 of \cite{AS}
associates to it a
$G$-equivariant  stable projective bundle over $M$
whose structure group  is $\PP U(\cH)$ with the norm topology (or the
compact-open topology),  satisfying some mild
local conditions. That is, there is a
 bundle of  projective spaces $ P$ with $G$-action,  mapping
$ P_x$ to $ P_{g\cdot x}$ by a projective isomorphism for any $x\in M$ and $g \in G$, and
satisfying
\begin{enumerate}
\item  $ P$ is  stable, i.e., $ P\cong  P \otimes L^2(G)$;
\item   each  point $x\in M$ with isotropy
group $G_x$ has a $G_x$-invariant neighbourhood $U_x$
such that there is an isomorphism of
bundles with $G_x$-action
\[
 P|_{U_x} \cong U_x\times \PP(\cH_x)
\]
for some projective Hilbert space $\PP(\cH_x)$  with $G_x$-action;
\item the transition functions between two trivializations are given by
maps
\[
U_x\cap U_y \longrightarrow Isom(\cH_x, \cH_y)
\]
which are continuous in the compact-open topology.
\end{enumerate}

As shown in \cite{MatSte}, given such a $G$-equivariant  stable projective bundle
$\cP$ over $M$, the lifting bundle gerbe (\cite{Mur}) associated to the corresponding
principal bundle and the central extension
$1\to U(1) \longrightarrow U(\cH) \longrightarrow \PP U(\cH) \to 0,$
is a $G$-equivariant bundle gerbe.
The following construction is, in a sense, a special case of this more general
approach.
With $LG$ being the smooth loop group
$C^\infty(S^1, G)$ and $\Omega G $ the based
loop group we have $LG = G \times \Omega G$.
There is a $G$-action on $\Omega G $ given by
conjugation.

\begin{proposition} \label{bg:G-equivariant}
Given an $LG$-manifold $\hat M$ with a  free
$\Omega G $-action such that the quotient map
\[
\pi: \hat{M} \rightarrow M:= \hat{M}/\Omega G
\]
defines a locally trivial principal $\Omega G$-bundle, then the lifting bundle gerbe over $M$
arising from a central extension
$1\to U(1) \longrightarrow \widehat{\Omega G} \longrightarrow \Omega G \to 1$
is a $G$-equivariant bundle gerbe over $M$.
\end{proposition}
\begin{proof} It is easy to see that
the quotient map $\pi: \hat{M}\to M$ is a $G$-equivariant surjective submersion.  The lifting
bundle gerbe is represented by the diagram
\[
   \xymatrix{
  \hat{g}^* \widehat{\Omega G} \ar[d] & \\
    \hat{M}^{[2]}\ar@< 2pt>[r]^{\pi_1} \ar@< -2pt>[r]_{\pi_2}
      &\hat{M}   \ar[d]\\
      &M
   }
   \]
   where $\hat{g}: \hat{M}^{[2]}\to \Omega G$ is determined by
   $x_2 = \hat{g} (x_1, x_2)\cdot x_1$ for $(x_1, x_2) \in \hat{M}^{[2]}$, and satisfies
   \[
    \hat{g} (x_2, x_3) \cdot \hat{g} (x_1, x_2) = \hat{g} (x_1, x_3)
    \]
    for $(x_1, x_2, x_3) \in \hat{M}^{[3]}$. The bundle gerbe product
    is given by
    \[
  m:   \bigl(\hat{g}^* \widehat{\Omega G}\bigr)_{(x_1, x_2) }
  \times \bigl(\hat{g}^* \widehat{\Omega G}\bigr)_{(x_2, x_3)}
  \to \bigl(\hat{g}^* \widehat{\Omega G}\bigr)_{(x_1, x_3) }
  \]
  which maps
  \[
 \Bigl( \bigl( x_1, x_2 , \hat{\gamma}_1\bigr)  ,
  \bigl( x_2, x_3 , \hat{\gamma}_2\bigr)\Bigr) \mapsto
    \bigl( x_1, x_3 , \hat{\gamma}_1\cdot \hat{\gamma}_2\bigr)
  \]
  for $\hat{\gamma}_1\in (\widehat{\Omega G})_{\hat{g}(x_1, x_2)}$ and
  $\hat{\gamma}_2\in (\widehat{\Omega G})_{\hat{g}(x_2, x_2)}$. Under the conjugation action
  of $G$ on  $\Omega G$, we see that the central extension
$  1\to U(1) \longrightarrow \widehat{\Omega G} \longrightarrow \Omega G \to 1$
is $G$-equivariant. This implies that
$\hat{g}^*\widehat{\Omega G}$ is $G$-equivariant. It remains to show that
the bundle gerbe product $m$ is $G$-equivariant. Using the fact that
the $\Omega G$-action on $\hat M$  is free and a direct calculation from
the definition of $\hat g$, we obtain, for $g\in G$
\[
\hat{g} (g\cdot x_1, g\cdot x_2) = g \cdot \hat {g}(x_1, x_2) \cdot g^{-1}.
\]
  From this equation we deduce the following
commutative diagram
\[
\xymatrix{
\Bigl( \bigl( x_1, x_2 , \hat{\gamma}_1\bigr)  ,
  \bigl( x_2, x_3 , \hat{\gamma}_2\bigr)\Bigr) \ar@{|->}[r]^m \ar@{|->}[d]^g
   & \bigl( x_1, x_3 , \hat{\gamma}_1  \hat{\gamma}_2\bigr)\ar@{|->}[d]^g \\
   \Bigl( \bigl (g  x_1, g  x_2 , Ad_g(\hat{\gamma}_1)\bigr)  ,
  \bigl( g  x_2, g  x_3 , Ad_g( \hat{\gamma}_2)\bigr)\Bigr) \ar@{|->}[r]
   & \bigl( g  x_1, g  x_3 ,Ad_g( \hat{\gamma}_1  \hat{\gamma}_2)\bigr).
}
   \]
i.e., $m$ is $G$-equivariant. Hence the lifting bundle
$\hat{g}^*\widehat{\Omega G}$ over $M$ is
a $G$-equivariant bundle gerbe.
    \end{proof}

Given any positive energy projective representation  of $\Omega G$
acting on $\cH$ of level
determined by the central extension of $\Omega G$, we see that
\[
\hat{M} \times_{\Omega G} \PP(\cH) \longrightarrow M
\]
is a $G$-equivariant stable projective bundle over $M$ whose
invariant (Cf. \cite{AS}) in $H^3_G(M, \ZZ)$ agrees
with the equivariant Dixmier-Douady class of the
lifting bundle gerbe defined in Proposition \ref{bg:G-equivariant}.

Given a $G$-equivariant bundle gerbe $\cG = (\cG, m; Y, M)$ over $M$, a
rank $n$ $G$-equivariant
 bundle gerbe module of $\cG$ is a bundle gerbe module $(\cE, \rho)$,
such that $\cE$ is a $G$-equivariant Hermitian vector bundle over $Y$,
and the bundle
gerbe action $\rho$ in (\ref{bg:module}) is $G$-equivariant.

\begin{definition}
   We call a generalized rank $n$ bundle gerbe $D$-brane $(Q, \mu)$
   of a  $G$-equivariant bundle gerbe $\cG$
   equivariant if $Q$ is a $G$-manifold and  $\mu$ is $G$-equivariant with respect to
   the conjugate action of $G$ on itself such that
   $\mu^*(\cG)$ admits a rank $n$ $G$-equivariant
 bundle gerbe module.
\end{definition}

Following \cite{GawRei} for the Wess-Zumino-Witten model on a group manifold
   $G$, the conjugacy classes of $G$ give so-called symmetric $D$-branes.
 We will see that they  provide many examples of
 rank $1$ $G$-equivariant bundle gerbe $D$-branes
   in $G$ and in fact that any  quasi-Hamiltonian $G$-manifold
  corresponding to a pre-quantizable Hamiltonian $LG$-manifold at level $k$
   is a generalized rank $1$ $G$-equivariant bundle gerbe $D$-brane.

   \section{Bundle gerbe $D$-branes from group-valued moment maps}

  Until the end of  Section 5, $G$ will denote
   a compact, connected and simply-connected
simple Lie group with Lie algebra $\g$.
  We fix  a smooth infinite dimensional model of
    $BG$ by embedding $G$ into $U(N)$ and letting $EG$ be the Stiefel manifold of
    $N$ orthonormal vectors in a separable complex Hilbert space.

Let $<\cdot, \cdot>$ be the  normalized invariant inner
product on $\g$  such that the highest co-root with respect to a basis of the root
system for  a fixed
maximal torus in $G$  has norm $2$.
   Then $k<\cdot, \cdot>$ defines an element
    in $$H^4(BG, \ZZ)\cong H^3(G, \ZZ) \cong \ZZ,$$ which
    in turn determines a central extension of $LG$ at level $k$ (\cite{PS}):
$$    1\to U(1) \longrightarrow \widehat{LG} \longrightarrow LG \to 1.$$

There is a technical issue, namely we need to complete the smooth loop
loop group $LG$ in an appropriate Sobolev norm for
the ensuing discussion. None of the constructions in the previous Section
are changed by using this completion.
Thus in the above exact sequence we
let $LG$ consist of maps of a fixed Sobolev class $L^2_{p}$ ($p>3/2$).
The based loop groups will continue to be
denoted by $\Omega G$.
    The Lie algebra of $LG$ is the space
    of maps $L\g = Map(S^1, \g)$ of Sobolev class $L^2_{p-1}$. Denote by
    \[
    L\g^* = \Omega^1(S^1, \g)
    \]
    whose elements are
    of Sobolev class $p-1$. Note that $L\g^*\subset (Lg)^*$, via
    the natural pairing of $L\g^*$ and $L\g$
    \[
    (a, \xi) = \int_{S^1} <a, \xi>.
    \]

     We can view $L\g^*= \Omega^1(S^1, \g)$  as the affine space of $L^2_{p-1}$-connections
      on the trivial bundle $S^1\times G$, with an $LG$-action
     by gauge transformations (the affine coadjoint action at level $k$):
     \ba\label{gauge:action}
     \gamma \cdot A = Ad_\gamma A - k\gamma^*  \bar{\theta}.
     \na
       Then there is a well-defined
     holonomy map
     \[
     Hol= Hol_1: \qquad L\g^* \longrightarrow G
     \]
     defined by solving the differential equation
     \[
     Hol_s(a)^{-1} \disp{\frac{\partial}{\partial t}} Hol_s(a) =k^{-1} a, \qquad
     Hol_0(a) = e
     \]
     where $s$ is the coordinate of $\RR$, and $S^1= \RR/ \ZZ$.
     The holonomy map $Hol$ is equivariant with respect to the
     evaluation homomorphism $LG \to G$, $\gamma \mapsto \gamma (1)$,
      and the conjugate action of $G$ on itself.

     We remark that the holonomy map $Hol: L\g^* \longrightarrow G$ also defines
     the universal $\Omega G$-principal bundle over $G$, and for
     $a \in L\g^*= \Omega^1(S^1, \g)$,
     the stabilizer of $a$ for the $LG$-action, denoted by $(LG)_a$,  is diffeomorphic
     to $G_{Hol(a)}$, the centralizer of  $Hol(a)$ in $G$.

  \subsection{Equivariant bundle gerbes over $G$}

Denote by $\theta, \bar{\theta} \in \Omega^1(G, \g)$ the left- and right-
   invariant Maurer-Cartan forms. In a faithful matrix representation $\rho$
      of $G$, $\theta = \rho^{-1}d\rho$ and $\bar\theta =  d\rho\rho^{-1}$.
      Let $\Theta_k\in \Omega^3(G)$ be the canonical
   closed bi-invariant 3-form on $G$:
   \[
   \Theta_k = \disp{\frac {k}{12} <\theta, [\theta, \theta]>
     =    \frac {k}{12} <\bar \theta, [\bar \theta, \bar\theta]>.}
     \]
Then $\Theta_k$ represents an integral de Rham cohomology class of $G$ in
$H^3(G, \RR)$ defined by $k \in H^4(BG, \ZZ) \cong \ZZ$.

The lifting bundle gerbe construction of  \cite{Mur}
starts from the universal $\Omega G$-principal  bundle
     $ Hol:  L\g^* \longrightarrow G$
     and the central extension
$1\to U(1) \longrightarrow \widehat{\Omega G} \longrightarrow \Omega G \to 1$
determined by the element $k \in H^4(BG, \ZZ) \cong \ZZ$.
   Then set $\cG_k = \hat{g}^*\widehat{\Omega G }$
     where $\hat{g}: (L\g^*)^{[2]} \to \Omega G$ is defined
   by $\xi_2 = \hat{g} (\xi_1, \xi_2) \cdot \xi_1$ for $(\xi_1, \xi_2) \in (L\g^*)^{[2]}$.

   \begin{proposition} \label{lift:G-equivariant}
   The lifting bundle gerbe $\cG_k$ is a $G$-equivariant bundle
   gerbe over $G$, whose equivariant Dixmier-Douady class is the class
     in $H^3_G(G, \ZZ)\cong \ZZ$ represented by     $\Theta_k$.
   \end{proposition}
   \begin{proof}
   Under the identification $L\g^* \cong \Omega^1(S^1, \g)$ determined by $k \in H^4(BG, \ZZ) \cong\ZZ$,
   the $LG$-action on $L\g^*$ makes the
   holonomy map
   \[
   Hol: L\g^* \longrightarrow
    L\g^*/\Omega G \cong
     G
   \]
   a $G$-equivariant principal $ \Omega G$-bundle with the conjugation action of
   $G$ on   $ \Omega G$. Then from Proposition \ref{bg:G-equivariant} and the
   discussion after the proof of  Proposition \ref{bg:G-equivariant}, we
   see that $\cG_k$ is a $G$-equivariant bundle
   gerbe over $G$, and the Dixmier-Douady class
   agrees with the non-equivariant Dixmier-Douady class of $\cG_k$
   under the isomorphisms
   \[
   H^3_G(G, \ZZ) \cong \ZZ \cong  H^3 (G, \ZZ),
   \]
   for any connected, compact, simply-connected simple Lie group $G$.
   \end{proof}

\subsection{Quasi-Hamiltonian $G$-spaces and Hamiltonian $LG$-spaces}
 We begin with a review from \cite{AAM} of  the
   definition of a group-valued moment map for a
   quasi-Hamiltonian $G$-space.

     \begin{definition}\label{quasi-hamiltonian}
      A {\bf quasi-Hamiltonian $G$-space} is a $G$-manifold with
     an invariant 2-form $\omega \in \Omega^2(M)^G$ and an equivariant
     map $\mu \in C^\infty(M, G)^G$ such that
     \begin{enumerate}
     \item The differential of $\omega$ satisfies $d\omega = \mu^*\Theta_k$.
     \item The map $\mu$ satisfies $\iota(v_\xi) = \disp{\frac k2} \mu^*<
     \theta + \bar \theta, \xi >,$ where $v_\xi$ is the fundamental vector
     field on $M$ generated by $\xi\in \g$.
     \item At each $x\in M$, the kernel of $\omega_x$ is given by
      \[
      ker \omega_x = \{ v_\xi| \ \xi \in ker(Ad_{\mu(x)+1})\}.
      \]
    The map $\mu$ is called the Lie group valued moment map of the
    quasi-Hamiltonian $G$-space $M$.
    \end{enumerate}
    \end{definition}

    Basic examples of quasi-Hamiltonian $G$-spaces are provided by conjugacy classes
    ${\mathcal C }\subset G$ as in \cite{AAM}, where the
    one-to-one correspondence between Hamiltonian loop group manifolds
    with proper moment map and quasi-Hamiltonian $G$-manifold is established.

 \begin{definition} A {\bf Hamiltonian $LG$-manifold at level $k$} is a triple $(\hat{M}, \hat{\omega}, \hat{\mu})$,
     consisting of a Banach manifold $\hat{M}$ with a smooth $LG$-action,
     an invariant weakly symplectic (that is, closed  and weakly nondegenerate)
     2-form $\hat{\omega}$ and an equivariant moment map $\hat{\mu}: \hat{M}
     \to  L\g^*$:
     \[
     \iota(v_\xi) \hat{\omega}= d \int_{S^1}k <\hat{\mu}, \xi>.
     \]
\end{definition}

\begin{remark}  It will be important later to observe
that a Hamiltonian $LG$-manifold at level $k$ is a Hamiltonian
      $\widehat{LG}$-manifold $\hat{M}$ with an
      $\widehat{LG}$-equivariant moment map
   $$\hat{\mu}: \hat{M}\longrightarrow  L\g^*= L\g^* \times \{k\} \hookrightarrow
     L\g^*  \oplus \RR, $$
     where $\widehat{LG}$ acts on $L\g^* \times \{k\}$ by the conjugation action. This
      conjugation action defines an affine coadjoint action of $LG$ at level $k$.
\end{remark}
     A Hamiltonian $LG$-manifold $(\hat{M}, \hat{\omega}, \hat{\mu})$ at level $k$
     is {\bf pre-quantizable}  if $\hat{M}$ has an $\widehat{LG}$-equivariant
     Hermitian line bundle
     $\cL \to   \hat{M}$,   with an invariant connection $\nabla$ whose
     curvature is given by $-2\pi i \hat{\omega}$ and
     \[
     2\pi k i < \hat{\mu}, \xi> =  Vert(\xi_\cL).
     \]
      Here  $\xi_\cL$ denotes the fundamental vector field
     on $\cL$ and $Vert: T\cL \to T\cL$ is the vertical projection defined by
     the connection $\nabla$. We call $(\cL, \nabla)$ the {\bf  pre-quantisation line bundle}
      for       $(\hat{M}, \hat{\omega}, \hat{\mu})$.

     Given  a Hamiltonian $LG$-manifold $(\hat{M}, \hat{\omega}, \hat{\mu})$ at level $k$
     with a proper moment map, then the $\Omega G$-action on $\hat{M}$ is free
     and  the quotient space $\hat{M}/\Omega G$ is a compact, smooth manifold of
     finite dimension.
     We define the holonomy manifold of $\hat{M}$
     as
     $     M= \hat{M}/\Omega G, $
         then  the following diagram commutes
     \ba
     \label{LG-action}
     \xymatrix{ \hat{M}\ar[d]_{\pi} \ar[r]^{\hat{\mu}} & L\g^* \ar[d]^{Hol}\\
     M \ar[r]^{\mu} &G.
     }
     \na
     There exists a unique invariant 2-form $\omega$ such
     that $(M, \omega, \mu)$ is  a quasi-Hamiltonian $G$-manifold with
     Lie group valued moment map $\mu$.

     Conversely, given  a quasi-Hamiltonian $G$-manifold
     $(M, \omega, \mu)$, there exists a unique
     Hamiltonian $LG$-manifold $(\hat{M}, \hat{\omega}, \hat{\mu})$ at level $k$ such that
     $M= \hat{M}/\Omega G$, and the commutative diagram  (\ref{LG-action})
     holds.     In fact,  $\hat{M}= M \times_G L\g^*$ is
     a principal $\Omega G$-bundle over $M$ and the $LG$-invariant weakly  symplectic  2-form
     is given by
     $$\hat{\omega}= \pi^* \omega + \hat{\mu} ^*\varpi$$
     where $\varpi$ is the following 2-form on $L\g^*$
     \[
     \varpi =\disp{\frac 12 \int_{S^1}  ds < Hol^*_s(\bar{\theta}), \frac
     {\partial}{\partial s} Hol^*_s(\bar{\theta})}>,
     \]
     satisfying $Hol^* \Theta_k = - d\varpi$ (For details, see Theorem 8.3 in \cite{AAM}).

      \begin{example} (Cf. Proposition 3.1 in  \cite{AAM}) Choose a maximal torus $T$
      in $G$ with its Lie algebra $\t$.
      The integral lattice $\Lambda^\vee_r \subset \t$ (the co-root lattice) is
      the kernel of the exponential map $exp: \t \to T$.
      Let $R$ and $R^\vee$ be the root system and the co-root
    system of $ G$. The root and co-root lattices $\Lambda_r
\subset  \t^*$ and $\Lambda_r^\vee\subset  \t $ are the lattices
spanned by   $R$ and $R^\vee$ with their $\ZZ$-basis given by simple roots and simple co-roots
\[
\Delta = \{ \a_1, \cdots, \a_n\}, \ \ and \ \ \Delta^\vee= \{
\a_1^\vee, \cdots, \a_n^\vee\}\] respectively, where $\a^\vee  =
\disp{\frac{2 \a}{<\a, \a>}}$.   The weight and co-weight lattices
$\Lambda_w \subset  \t^*$ and $\Lambda_w^\vee\subset  \t $ are the
lattices dual to $\Lambda_r^\vee$ and  $\Lambda_r$.
    Then   affine coadjoint
      $LG$-orbits at level $k$ are labelled by
      \[
      \cU_k = \{ \lambda \in \t^* | <\lambda, \a_j^\vee > \geq 0 \text{ for any simple
      co-root $\a_j$}, <\lambda, \vartheta> \leq k\}
      \]
      where $\vartheta$ is the highest root in $R$ with respect to $\Delta$.
       We denote by $\cO_\lambda$ the affine  coadjoint  $LG$-orbit  at level $k$ through
        $\lambda \in \cU_k$.       The conjugacy classes  in $G$ are labelled
      by elements
      \[exp(\disp{\frac{2\pi i\lambda}{k}})
      \]
       under  $\g\cong \g^*$ defined by $k<\cdot, \cdot>$.
     A conjugacy class $\cC_\lambda$ (for $\lambda \in \cU_k $) in $G$
      \[
      \{g= g_0\cdot exp(\disp{\frac{2\pi i\lambda}{k}}) \cdot g^{-1}_0 | g_0 \in G\}
      \]
      has a canonical 2-form
      $$\omega (g)  = \disp{\frac k2} < \theta, (1-Ad_g)^{-1} (\theta)>  $$
      such that $d \omega = \Theta_k |_{\cC_\lambda}$ and   $(\cC, \omega) $ is a
      quasi-Hamiltonian   $G$-manifold with moment map the embedding $\cC \hookrightarrow G$.
       The corresponding Hamiltonian $LG$-manifold at level $k$ is
      given by   the   affine coadjoint  $LG$-orbit $\cO_\lambda$.
      Then $ \cO_\lambda$ as a Hamiltonian $LG$-manifold at level $k$
      is pre-quantizable if and only if
      $$\lambda \in \Lambda_k^*:=\Lambda_w \cap \cU_k,$$
  whose elements are called the dominant weights  at level $k$. The pre-quantization
  line bundle over $\cO_\lambda$ is given by
  \[ \widehat{LG}\times_{\widehat{LG}_{\lambda}} \CC_{(*\lambda, 1)}
  \]
  where $\widehat{LG}_{\lambda}$ acts on $\CC_{(*\lambda, 1)}$ with weight $(*\lambda, 1)$,
  $*\lambda$ is the dominant weight of the irreducible representation of $G$
   complex conjugate to the one with weight $\lambda$..
   The geometric quantization
  on $\cO_\lambda$   by the Borel-Weil construction  as in \cite{PS}  gives rise to the irreducible
  positive energy representation of $\widehat{LG}$ with the highest
  weight $(\lambda, k)$.
            \end{example}

    \subsection{Equivariant bundle gerbe modules}

     Our main theorem in this Section is the following
     existence result for generalized $G$-equivariant bundle
   gerbe modules in terms of pre-quantizable Hamiltonian $LG$-manifolds at level $k$.

     \begin{theorem}\label{bg:module:rank1}
     Given a quasi-Hamiltonian $G$-manifold $(M, \omega, \mu)$ such that
     the corresponding Hamiltonian $LG$-manifold $(\hat{M}, \hat{\omega}, \hat{\mu})$ at
     level $k$ is pre-quantizable and the moment map $\hat{\mu}$ is proper, then the
     pull-back of the bundle gerbe $\cG_k$ over $G$, $\mu^*\cG_k$,
     admits a canonical $G$-equivariant trivialization
     \[
     \xymatrix{
  \mu^*\cG_k\cong \delta (\cL_{\hat M})\ar[d] \quad \qquad & \\
    \hat{M}^{[2]}\ar@< 2pt>[r]^{\pi_1} \ar@< -2pt>[r]_{\pi_2}
      &\hat{M}   \ar[d]\\
      &M
   }
   \]
   where $\cL_{\hat M}$ is the pre-quantization line bundle over $\hat M$, and
   $ \delta (\cL_{\hat M})= \pi_2^*\cL_{\hat M}\otimes \pi_1^*\cL_{\hat M}^{-1}$.
   \end{theorem}
   \begin{proof}
   The pull-back of the bundle gerbe $\cG_k$ is determined
   by the following diagram:
   \[
     \xymatrix{
  \mu^*\cG_k\ar[d] & \\
    \hat{M}^{[2]}\ar@< 2pt>[r]^{\pi_1} \ar@< -2pt>[r]_{\pi_2}
      &\hat{M}   \ar[d]\\
      &M
   }
   \]
   Specifically  $ \mu^*\cG_k$ is the pullback
to  $\hat{M}^{[2]}$
of the $U(1)$ bundle determined by the central extension
   $U(1) \to \widehat{\Omega G}\to \Omega G$
   corresponding to $\Theta_k$ under the map
$\hat{g_M}: \hat{M}^{[2]} \to \Omega G$ defined
   by $x_2 = \hat{g_M} (x_1, x_2) \cdot x_1$ for $(x_1, x_2) \in \hat{M}^{[2]}$. For a
   pre-quantizable Hamiltonian $LG$-manifold $(\hat{M}, \hat{\omega}, \hat{\mu})$ at level $k$,
   the  pre-quantization line bundle $\cL_{\hat M}$ carries an
$\widehat{\Omega G}$-action
   such that the following diagram commutes:
   \[ \xymatrix{
     \widehat{\Omega G} \times \cL_{\hat M}  \ar[r] \ar[d] & \cL_{\hat M}\ar[d]  \\
     \Omega G \times \hat{M}  \ar[r]  & \hat{M}.
      }
        \]
   This implies that, for $(x_1, x_2) \in \hat{M}^{[2]}$
   \[
   (\mu^*\cG_k)_{(x_1, x_2)} \otimes (\cL_{\hat M})_{x_1} \cong (\cL_{\hat M})_{x_2}.
   \]
   The associativity of the $\widehat{\Omega G}$-action ensures that
    $\cL_{\hat M}$ is a rank one bundle gerbe module of $\mu^*\cG_k$.
   As a rank one bundle gerbe module of
   $\mu^*\cG_k$, we know that
   $$\mu^*\cG_k \cong\pi_2^*\cL_{\hat M}\otimes \pi_1^*\cL_{\hat M}^{-1}. $$

   As the pre-quantization line bundle  $\cL_{\hat M}$ is actually an
$\widehat{LG}$-equivariant
   line bundle, we immediately know that
   the rank one bundle gerbe module $(\cL_{\hat M}, \hat{M})$ is $G$-equivariant
   in the sense that $\cL_{\hat M}$ is a $G$-equivariant line
   bundle over $\hat{M}$ and the bundle gerbe action
   \[
   \mu^*\cG_k \otimes\pi_1^* \cL_{\hat M} \cong \pi_2^* \cL_{\hat M}
   \]
   is $G$-equivariant.
      \end{proof}

     \begin{remark} In the  set-up of  differentiable
 stacks and their presenting Lie groupoids, a more general
 moment map theory is developed in \cite{Xu}. Results
  analogous to Theorem \ref{bg:module:rank1} in terms of pre-quantizations
  of quasi-symplectic groupoids and the compatible pre-quantizations
  of  their quasi-Hamiltonian spaces   are also discussed in \cite{LauXu}
  for non-equivariant cases.
 \end{remark}

     From this theorem, we can deduce easily the following existence result
for equivariant bundle gerbe $D$-branes of the bundle gerbe
       $\cG_k$ over $G$.

      \begin{corollary} \label{brane:exist}
      Given a quasi-Hamiltonian $G$-manifold $(M, \omega, \mu)$ such that
      the corresponding Hamiltonian $LG$-manifold $(\hat{M}, \hat{\omega}, \hat{\mu})$
     at level $k$ is pre-quantizable and proper, then  $(M, \omega, \mu)$ is a generalized
      rank one $G$-equivariant bundle
      gerbe $D$-brane of $\cG_k$.
      \end{corollary}

    For the quasi-Hamiltonian $G$-manifolds from conjugacy classes
 $\cC_\lambda$ in $G$,
 we obtain the corresponding symmetric $D$-brane in $G$, which
 is pre-quantizable if and only if $\lambda$ is a dominant weight at level $k$.

\begin{remark}
Relax the pre-quantizable condition in Corollary \ref{brane:exist}
on   $(\hat{M}, \hat{\omega}, \hat{\mu})$
 to allow   $\hat{M}$ to have an $\hat{LG}$-equivariant
     Hermitian vector  bundle   $\cE \to   \hat{M}$ of rank $n$,
      with an invariant connection $\nabla$ whose
     curvature is given by $-2\pi i \hat{\omega}\otimes Id$. Then the same proof
     implies that there exists a $G$-equivariant bundle gerbe action
     \[
     \mu^*\cG_k \otimes\pi_1^* \cE \longrightarrow \pi_2^* \cE,
     \]
     that is to say,
         $(M, \omega, \mu)$ is a generalized
      $G$-equivariant bundle   gerbe $D$-brane of rank $n$.
      \end{remark}

  \section{Moduli spaces of flat connections on Riemann surfaces}
  \label{section:moduli}

  In this section, we recall a particular class of quasi-Hamiltonian
$G$-manifolds,
  given by the moduli spaces of flat connections on Riemann surfaces with
boundaries.

 Denote  by $\cA_{S^1}$  the space of
 $L^2_p$ $G$-connections
 on the trivial principal $G$-bundle over $S^1$. The holonomy map
$ Hol:  \cA_{S^1}\longrightarrow G$
  defines  a principal $\Omega G$-bundle over $G$, where $\Omega G$ is the
  based loop group, identified with the based gauge transformation
  group  under a choice of parametrization of $S^1$.  With a fixed trivial
  connection, we can identify $\cA_{S^1}$ with $\Omega^1(S^1, \g)$. Using the
  isomorphism $\g \cong \g^*$ defined by $<, >$, we have
  $\cA_{S^1} \cong L\g^*.$ We know that $Hol:  \cA_{S^1}\rightarrow G$ agrees
  with the universal
  $\Omega G$-bundle over $G$ constructed in proposition 3.2.
  The bundle gerbe $\cG_k$ over $G$ is the lifting bundle
  gerbe (Cf. \cite{Mur}) associated to the principal $\Omega G$-bundle
  $\cA_{S^1}\to G$:
  \ba
  \label{bg:G}
  \xymatrix{
  \cG_k  \ar[d]& \\
     \cA_{S^1}^{[2]}\ar@< 2pt>[r]^{\pi_1} \ar@< -2pt>[r]_{\pi_2}
     &  \cA_{S^1}  \ar[d] \\
     &G}
  \na
  such that $\cG_k = \hat{g}^*\bigl(\widehat{\Omega G}\bigr)$,
  where $\widehat{\Omega G}$ is  the central extension
  $U(1) \to \widehat{\Omega G}\to \Omega G$
  determined
  by $\phi\in H^4(BG, \ZZ)$ (Cf. \cite{PS}), and
  $\hat{g}:\cA_{S^1}^{[2]}\to\Omega G$ is defined by $A_2 = A_1\cdot \hat{g}(A_1, A_2)$
  for $ (A_1, A_2)\in \cA_{S^1}^{[2]}$.  The full gauge group (identified with $LG$) action and
  Proposition  \ref{bg:G-equivariant} tell us that $\cG_k$ is
  a $G$-equivariant bundle gerbe over $G$ with equivariant Dixmier-Douady
  class $k \in \ZZ \cong H^3_G(G, \ZZ)$.

  The classifying  map for the universal $\Omega G$-bundle over $G$,  a
  homotopy equivalence between   $\Omega (BG)$ and $B(\Omega G)$ and the evaluation map
  from $S^1 \times \Omega(B G)$ to $ BG$
  define   a homotopy class of maps, formally denoted by  $[ev]$:
  \[
  [ev]: S^1 \times G \sim  S^1 \times B\Omega G \sim S^1 \times \Omega(B G) \to  BG,
  \]
  such that the Dixmier-Douady class of $\cG_k$ is given by the cohomology class
  $$[\omega_k]= \disp{\int_{S^1}\circ [ev] } (\phi), $$ where $\phi \in H^4(BG, \ZZ)$
  is defined by $<, >$.

  Now, given a Riemann surface with one boundary component
  which is {\sl pointed} by fixing a base point on the boundary, denote
  by $\cM_{\Sigma}$
  \[
  \disp{\frac{ \{\text{flat $L^2_{p-\frac 12}$ $G$-connections on $\Sigma \times G$}\}}
 {\{g: L^2_{p+\frac 12}(\Sigma, G)|  g (\text{base point}) = Id \in G\}}},
 \]
   the based moduli space of flat $G$-connections on $\Sigma$
  (that is, the space of  flat $G$-connections on $\Sigma\times G$ modulo
  gauge transformations which are the identity at the base point).
  Note that the based moduli space $\cM_\Sigma$ is a finite
  dimensional smooth manifold and
  the boundary holonomy map defines a group valued moment map (Cf. \cite{AAM})
  $ \mu_\Sigma: \cM_{\Sigma} \longrightarrow G$.
   There is a canonical principal  $G$-bundle with connection
     over $\Sigma \times  \cM_{\Sigma}$ given by choosing a
  classifying map
  \ba\label{Ev}
  Ev: \qquad \Sigma \times  \cM_{\Sigma} \longrightarrow BG,
  \na
  such that $ev$, the  restriction of $Ev$
  to $\partial \Sigma \times  \cM_{\Sigma}  = S^1 \times \cM_{\Sigma}$,
    represents a map  in the homotopy class of maps:
  \[
  \xymatrix{
  S^1 \times \cM_{\Sigma} \ar[r]^{Id \times \mu_\Sigma} & S^1 \times
  G \ar[r]^{[ev]} & BG.
  }
  \]

 Now we can represent $\phi$ by a differential form $\Phi (\disp{\frac{i}{2\pi}} F_\AA)$
 on $BG$, where $\Phi$ is the corresponding
$G$-invariant degree two polynomial on the Lie algebra $\g$ of $G$,
determined by the  inner product $< \cdot, \cdot, >$ on $\g$.
Then from direct calculation, we see  that
  \[
  d (\disp{\int_\Sigma} Ev^* \Phi (\disp{\frac{i}{2\pi}} F_\AA))
  = \mu_\Sigma^*\disp{\int_{S^1}} ev^* \Phi (\disp{\frac{i}{2\pi}} F_\AA).
  \]
  Hence $\mu_\Sigma^*\omega_k$ is exact so the Dixmier-Douady
  class of
  the  pullback bundle gerbe  $\mu^*_\Sigma\cG_k$  over $\cM_\Sigma$ is
 trivial.

  \begin{proposition}\label{moduli:gb-module} The quasi-Hamiltonian
  $G$-space $(\cM_{\Sigma}, \mu_\Sigma)$ determines a unique Hamiltonian
  $LG$-space at level $k$ which is diffeomorphic to
  $$\mu^*_\Sigma \cA_{S^1}  =\cM_{\Sigma}\times_G \cA_{S^1} \cong \hat{\cM}_\Sigma, $$
  with $\hat{\cM}_\Sigma$ is the moduli space
    of flat connections modulo gauge transformations which are the identity
    on the boundary.
  Moreover the Hamiltonian $LG$-manifold $\hat{\cM}_\Sigma$ is pre-quantizable and
  admits a proper moment map, hence $(\cM_{\Sigma}, \mu_\Sigma)$ is a generalized rank 1
  $G$-equivariant bundle  gerbe $D$-brane of $\cG_k$.
  \end{proposition}
   \begin{proof}
 By results in  \cite{AB} and \cite{Don},  $\hat{\cM}_\Sigma$ is an
 infinite dimensional symplectic manifold admitting a residual Hamiltonian
 action of  $L G$ at level $k$, whose moment map is given by the pullback of connections to the boundary
 \[
 \hat{\mu}_\Sigma: \hat{\cM}_\Sigma \longrightarrow \cA_{S^1}\cong L\g^*.
 \]
  The induced $\Omega G$-action
 is free on $\hat{\cM}_\Sigma$
 such that the quotient map $\hat{\cM}_\Sigma\to\cM_\Sigma $
 is the induced principal $\Omega G$-bundle
 and  the following diagram is commutative:
 \ba\label{diagram:1}
 \xymatrix{ \hat{\cM}_\Sigma\ar[d]_{\pi} \ar[r]^{\hat{\mu}_\Sigma} & L\g^* \ar[d]^{Hol}\\
     \cM_\Sigma \ar[r]^{\mu_\Sigma} &G.
     }
  \na
  which confirms  $\mu^*_\Sigma \cA_{S^1} \cong \hat{\cM}_\Sigma$, and the moment
  map $\hat{\mu}_\Sigma$ is proper.

  Now we give a   construction of a pre-quantization line bundle
  following  \cite{MeiWoo} and \cite{AB}.
 Denote by $\cA^{flat}_\Sigma$ the space of flat $G$-connections on $\Sigma \times G$, and
 denote by $\cG_0(\Sigma)$ and $\cG_0(\partial \Sigma)$ the
 based gauge transformation groups on $\Sigma$ and $\partial \Sigma$ respectively.
 Let $\cG (\Sigma, \partial\Sigma)$ be the kernel of the restriction map to the boundary
 \ba\label{partial}
 \partial: \qquad \cG_0 (\Sigma )\longrightarrow \cG_0 (\partial \Sigma )
 \cong \Omega G
 \na
 then $\cG (\Sigma, \partial\Sigma)$ consists of those gauge transformations
 which are the identity on the whole boundary.
 Since $G$ is simply connected,  we have the following exact sequence
 \[
 1\to \cG (\Sigma, \partial\Sigma)\longrightarrow \cG_0 (\Sigma)
 \longrightarrow \cG_0 (\partial \Sigma )\to 1,
 \]
 and the principal $\Omega G$-bundle
$ \hat{\cM}_\Sigma
  \longrightarrow\cM_\Sigma$
 is induced by the residual action of the gauge group $\cG_0 (\Sigma)$.

 The pullback of the central extension
$ 1\to U(1) \longrightarrow \widehat{\Omega G} \longrightarrow \Omega G  \to 1$
  under the map $\partial$ (\ref{partial})
 defines a central  extension $\widehat{\cG_0}(\Sigma)$ of $\cG_0(\Sigma)$ whose
 2-cocycle is given by
 \ba\label{2-cocyle}
 c(g_1, g_2) = exp\bigl(2\pi i \disp{\int_\Sigma}< g_1^{-1} dg_1,
 dg_2 g_2^{-1}> \bigr),
 \na
 It is known that this extension has
 a canonical trivialisation  over $\cG (\Sigma, \partial\Sigma)\subset \cG_0(\Sigma) $.

 The pre-quantization line bundle over $\cA^{flat}_\Sigma$ is given
 by the trivial line bundle $\cA^{flat}_\Sigma \times \CC$ with connection
 1-form
 \ba\label{connection}
 \begin{array}{cc}
 \theta_A: &T_A\cA^{flat}_\Sigma \cong \{ \a\in \Omega^1(\Sigma, \g)|
 d\a = 0 \}\to  \RR\\[2mm]
  &\a  \mapsto  \disp{\int_\Sigma}<\a , A >.
 \end{array}
 \na
 Here $\Omega^1(\Sigma, \g)$ is the space of Lie algebra $\g$ valued
 1-form on $\Sigma$.
  This pre-quantization line bundle admits a connection-preserving action
 of $\widehat{\cG_0}(\Sigma)$ via the local action
 \[
 (g, z)\cdot (A, w) = (g\cdot A, exp\bigl(-2\pi i \disp{\int_\Sigma}< g^{-1} dg,
 A> \bigr) zw),
 \]
  whose quotient under the $\cG (\Sigma, \partial\Sigma)$-action
  \[
 \cL_\Sigma =  (\cA^{flat}_\Sigma \times \CC )/\cG (\Sigma, \partial\Sigma)
  \]
  is the  pre-quantization line bundle over $\hat{\cM}_\Sigma$.

  We claim that $\mu_\Sigma:\cM_\Sigma \to G$ is a generalized rank 1 bundle
  gerbe $D$-brane, with the canonical
 trivialisation of $\mu^*_\Sigma\cG_k$  given by
 \ba\label{bgmodule:cM}
 \xymatrix{
  \mu^*_\Sigma\cG_k\cong \delta (\cL_{\Sigma})\ar[d] \quad \qquad & \\
    \hat{\cM}_\Sigma^{[2]}\ar@< 2pt>[r]^{\pi_1} \ar@< -2pt>[r]_{\pi_2}
      &\hat{\cM}_\Sigma  \ar[d]\\
      &\cM_\Sigma
   }
  \na
 where $ \cL_\Sigma$ is the pre-quantization line bundle over $\hat{\cM_\Sigma}$
  in \cite{MeiWoo}\cite{Wit}.

   As $\cL_\Sigma$ carries an action of
 $LG$, it is straight forward to show that
 $ \cL_\Sigma$ is a  $G$-equivariant
 bundle gerbe module of $\mu^*_\Sigma \cG_k$ (see the proof of Theorem
  \ref{bg:module:rank1}),
 therefore, we have shown that
  $\mu_\Sigma^* \cG_k \cong \delta (\cL_\Sigma)$ as in  (\ref{bgmodule:cM}).
  The connection $\theta$ (\ref{connection}) on $\cA^{flat}_\Sigma \times \CC$
  descends to a bundle gerbe module connection on $\cL_\Sigma$.
Hence,  $(\cM_{\Sigma}, \mu_\Sigma)$ is a generalized rank 1
  $G$-equivariant bundle  gerbe $D$-brane of $\cG_k$.
\end{proof}

\subsection{Relationship with Chern-Simons.}

In this subsection we summarise some
observations about the present situation
which may be deduced from \cite{CJMSW}. In that paper
 the universal Chern-Simons bundle
  2-gerbe $\cQ_\phi$ associated to
$\phi \in H^4(BG, \ZZ)$ is defined to be a bundle 2-gerbe $(\cQ_\phi, EG^{[2]}; EG, BG)$
with connection
 illustrated by the following diagram:
  \ba\label{CS:b2g}
  \xymatrix{
  & \cG_{\tau(\phi)} \ar@2{->}[d]\\
 \cQ_\phi \ar@2{->}[d]& G  \\
  EG^{[2]}\ar[ur]^{\hat{g}}\ar@< 2pt>[r]^{\pi_1} \ar@< -2pt>[r]_{\pi_2}
  &  EG  \ar[d]_{\pi} \\
   & BG
  }
  \na
The way to read this diagram is that $\cQ_{\phi}$ is obtained
as the pull-back of  a multiplicative bundle
  gerbe $\cG_{\tau(\phi)}$ over $G$,
with connection and curving, whose bundle gerbe curvature
$\tau (\phi)\in H^3(G, \ZZ)$ is determined by
$\phi$ and
where $\tau:H^4(BG, \ZZ) \to H^3(G, \ZZ)$ is the usual transgression map.

The technicalities in \cite{CJMSW} are handled by
recognising that transformations
between stable isomorphisms of bundle 1-gerbes provide $2$-morphisms making
the category $\mathbf{BGrb}_M$ of bundle 1-gerbes over a manifold $M$
and stable isomorphisms between bundle 1-gerbes into a bi-category.
The space of 2-morphisms between
two stable isomorphisms is in one-to-one correspondence with
the space of line bundles over $M$.

 We also recall from \cite{CJMSW} the definition of a multiplicative bundle
  gerbe on a compact semi-simple Lie group $G$.
  Let $BG_{\bullet}$ be the following simplicial manifold
  \[
  BG_{\bullet} =\{BG_n =  G \times \cdots \times G \ \text{(n copies)} \}
  \]
   (where $n=0, 1, 2,
  \cdots$), endowed with  face operators $\pi_i: G^{n+1} \to G^n$,
  ($i= 0, 1, \cdots, n+1$)
  \[
  \pi_{i}(g_{0},\ldots,g_{n}) = \begin{cases}
                                  (g_{1},\ldots,g_{n}), & i = 0, \\
                                  (g_{1},\ldots,g_{i-1}g_{i},g_{i+1},
                                    \ldots,g_{n}), & 1\leq i\leq n,
  \\
                                   (g_{0},\ldots,g_{n-1}), & i = n+1.
                                  \end{cases}
  \]
  In particular, the face operators from $G\times G \to G$
  consist of $\pi_0(g_1, g_2) = g_2$, $\pi_1(g_1, g_2) = g_1g_2$ and $\pi_2(g_1, g_2) = g_1$ for
  $(g_1, g_2) \in G \times G$.

   The face operator $\pi_i: G^{n+1} \to G^n$ defines a bi-functor
\[
\pi_i^*:   \mathbf{BGrb}_{G^{n}}  \longrightarrow
\mathbf{BGrb}_{G^{n+1}}
\]
sending objects, stable isomorphisms and 2-morphisms to
the pull-backs by $\pi_i$.

  \begin{definition}\label{multiplicative} (Cf. \cite{CJMSW})
   A multiplicative bundle
  gerbe on $G$ is a   bundle gerbe $\cG$ over $G$ together with a
  stable isomorphism  $$m: \pi_0^*\cG \otimes \pi_2^* \cG  \to \pi_1^*\cG$$
  over $G\times G$, where $\pi_i^*\cG$ is the pull-back bundle gerbe over $G\times G$,
   such that,   the stable isomorphism  $m$
  is associative up to a 2-morphism in $\mathbf{BGrb}_{G\times G \times G}$:
  \[
  \varphi: \pi_2^*m \circ ( \pi_0^*m \otimes Id)\Longrightarrow
  \pi_1^*m \circ (Id \otimes \pi_3^*m),
  \]
   for which the corresponding
  line bundle $\cL_\varphi$ over $G\times G \times G$ is trivial.
Moreover, there is a canonical isomorphism between two trivial line bundles
over $G^4$ with their induced trivializing sections:
$$\pi^*_1\cL_\varphi \otimes \pi^*_3\cL_\varphi \otimes \pi^*_5\cL_\varphi\cong
 \pi^*_2\cL_\varphi \otimes \pi^*_4\cL_\varphi.$$
   \end{definition}

The main result of \cite{CJMSW}
is that a bundle gerbe $\cG$ over $G$ is multiplicative if and only if
  the corresponding Dixmier-Douady class lies in the image
  of the transgression map $\tau: H^4(BG, \ZZ) \rightarrow H^3(G, \ZZ).$
The relation between the Chern-Simons bundle gerbe
and the moduli space of flat $G$-connections
 is given by the following proposition.

 \begin{proposition} The transgression of our universal Chern-Simons bundle
  2-gerbe $\cQ_\phi$ to the moduli space of flat $G$-connections on
  a closed Riemann  surface is the Chern-Simons line bundle over the moduli
space.
   \end{proposition}
 \begin{proof}
 Given a closed  Riemann surface $\Sigma$ with a base point we
 cut $\Sigma$ along a separating simple curve through the base point so that
$\Sigma = \Sigma_1\cup_{S^1}\Sigma_2$
and $\Sigma_i$ is a Riemann surface with one pointed boundary.
It is easy to see that  the based moduli space of flat $G$-connections
on $\Sigma\times G$ is given by the fiber product of the group valued moment maps for
 $\cM_{\Sigma_1}$ and $\cM_{\Sigma_2}$
\ba
\begin{array}{lll}
\cM_{\Sigma} &\cong & \cM_{\Sigma_1}\times_G \cM_{\Sigma_2}\\[2mm]
&\cong& \bigl( \hat{\cM}_{\Sigma_1}\times\hat{\cM}_{\Sigma_2}
\bigr) // \Omega G\\[2mm]
&\cong &\bigl( \hat{\cM}_{\Sigma_1}\times_{\cA_{S^1}}\hat{\cM}_{\Sigma_2}
\bigr) / \Omega G,
\end{array}
\na
with the induced map $\mu_\Sigma: \cM_{\Sigma}\to G$. Here we use the notation
 $``//"$ for the symplectic reduction  by the diagonal $\Omega G$-action
  with respect to the
 moment map $\hat{\mu}_{\Sigma_1} -  \hat{\mu}_{\Sigma_2}$,
 as $0$ is a regular value (\cite{AB}),
 \[ (\hat{\mu}_{\Sigma_1} -  \hat{\mu}_{\Sigma_2})^{-1} (0)
 = \hat{\cM}_{\Sigma_1}\times_{\cA_{S^1}}\hat{\cM}_{\Sigma_2},
 \]
   and the action of $\Omega G$ is
 free.

The pull-back bundle gerbe $\mu_\Sigma^*\cG_k$ over $\cM_{\Sigma}$
from $\cG_k$ over $G$ now has two trivialisations from the canonical
trivialisations of the pull-back bundle gerbes over
$\cM_{\Sigma_1}$ and $\cM_{\Sigma_2}$ (see Proposition \ref{moduli:gb-module}).
These two trivialisations give a line
bundle  $\cL_\Sigma$ over $\cM_{\Sigma}$  which we refer to as the
transgression of $\cQ_\phi$. In \cite {CJMSW}, there is a natural
bundle 2-gerbe connection on $\cQ_\phi$. The Chern-Simons bundle 2-gerbe connection
induces canonical bundle gerbe module connections on  the bundle gerbe  modules
$\cL_{\Sigma_1}$ and $\cL_{\Sigma_2}$.  These  bundle gerbe module connections
define a canonical connection
on the line bundle  $\cL_\Sigma$.  The curvature of this
canonical connection is given by
\[
\disp{\int_{\Sigma_1}} Ev_1^* \Phi (\disp{\frac{i}{2\pi}} F_{\AA_1})
- \disp{\int_{-\Sigma_2}} Ev_2^* \Phi (\disp{\frac{i}{2\pi}} F_{\AA_2})
= \disp{\int_{\Sigma} }Ev^* \Phi (\disp{\frac{i}{2\pi}} F_\AA),\]
where $Ev$ is the classifying map for the canonical
$G$-bundle over $\Sigma\times \cM_\Sigma$ with connection $Ev^*\AA$, and
$\Phi$ is the corresponding
$G$-invariant degree two polynomial on the Lie algebra $\g$ of $G$.
This agrees with the curvature formula in \cite{AB}\cite{RSW}
for the  Chern-Simons line bundle over $\cM_\Sigma$.
\end{proof}

\subsection{Recasting the Segal-Witten reciprocity law}

   We now   interpret  the Segal-Witten reciprocity law   (Cf. \cite{BryMcL}) from the
   viewpoint   of bundle gerbes over $G$ and their  bundle gerbe $D$-branes.

  Let $G_\CC$ be the complexification of a connected, compact and semi-simple
  Lie group (not necessarily simply-connected) $G$. Let $LG_\CC$
  denote the smooth loop group.  A central extension of $LG_\CC$ by $\CC^*$
  \[
  1\to \CC^* \longrightarrow \widehat{LG_\CC} \longrightarrow LG_\CC \to 1
  \]
  has the  reciprocity property if, for an extended Riemann surface $\Sigma$ whose
  boundary $\partial \Sigma$ is a disjoint union of
  parametrized circles, the extension of
  $C^\infty( \partial \Sigma, G_\CC)$ induced by the Baer product of the extension of boundary
  components
  \[1\to \CC^* \longrightarrow \widehat{C^\infty(\partial\Sigma, G_\CC)} \longrightarrow
    C^\infty(\partial\Sigma,  G_\CC) \to 1
    \]
  {\bf splits canonically} over the subgroup of holomorphic
  maps (denoted by   $Hol (\Sigma, G_\CC)$) from $\Sigma$ to $G_\CC$.

   We use
  $s_\Sigma:Hol (\Sigma, G_\CC) \to\widehat{Hol (\Sigma, G_\CC)} $ to denote the canonical section of the  splitting of the
induced extension
  \[
  1\to \CC^* \longrightarrow \widehat{Hol (\Sigma, G_\CC)} \longrightarrow
    Hol (\Sigma, G_\CC) \to 1.
    \]

  Given a  central extension $\widehat{LG_\CC} $ with
  the  reciprocity property,  $\widehat{LG_\CC} $
satisfies the glueing property
  if whenever
an extended Riemann surface is obtained by glueing two extended
  Riemann surfaces $\Sigma_1$ and $\Sigma_2$ along some
  boundary components with the obvious
  restriction maps $$\rho_i: Hol (\Sigma, G_\CC) \longrightarrow
  Hol (\Sigma_i, G_\CC), $$
  then there is a canonical isomorphism between
  $ \widehat{Hol (\Sigma, G_\CC)}$ and $\rho_1^*  Hol (\Sigma_1, G_\CC)\otimes
  \rho_2^*  Hol (\Sigma_2, G_\CC)$ carrying the section
  $s_\Sigma$ to $\rho_1^*s_{\Sigma_1} \otimes \rho_2^*s_{\Sigma_2}$.

  The  Segal-Witten reciprocity law claims
  that an extension $\widehat{Hol (\Sigma, G_\CC)}$ satisfies the  reciprocity
  and glueing properties if the characteristic class of the extension lies
  in the image of the transgression map
  $\tau: H^4(BG, \ZZ) \longrightarrow H^3(G, \ZZ).$
   In \cite{BryMcL}, the converse of the
  Segal-Witten reciprocity law is established:  any extension of $LG_\CC$
   satisfying the reciprocity
  and glueing properties must lie in the image of $\tau$.

  In the light of the properties of multiplicative bundle gerbes,
  their bundle gerbe modules and their
  relationship with the Chern-Simons bundle 2-gerbes, we can recast the
  Segal-Witten reciprocity law as in the following proposition for a simply-connected
  simple Lie group $G$.

 \begin{proposition}
 \label{reciprocity}
 For an extended Riemann surface $\Sigma= \Sigma_{g, n}$ of genus $g$
   with  $n$ pointed and parametrized boundary
  components, the transgression of the Chern-Simons bundle 2-gerbe $\cQ_\phi$
  (\ref{CS:b2g}) provides a canonical $G$-equivariant trivialization of the
  pull-back equivariant bundle gerbe associated to the
  group-valued moment map
  $\mu_\Sigma:\cM_\Sigma \longrightarrow G ^n.$
  \end{proposition}
 \begin{proof}
  Given Riemann surface $\Sigma=\Sigma_{g, n}$ of genus $g$
   with $n$ pointed boundary components
  $\bigsqcup_{i=1}^n S_i$,  the boundary holonomy map defines the group
  valued moment map for the based moduli space $\cM_\Sigma$,
 $\mu_\Sigma:\cM_\Sigma \longrightarrow G ^n.$
 We can identify as before $\mu_\Sigma^*(\cA_{S^1}^n)$
 with
 \[
 \hat{\cM}_\Sigma = \disp{\frac{ \{\text{flat $G$-connections on $\Sigma \times G$}\}}
 {\{g: \Sigma \to G| \quad g|_{\partial \Sigma} = Id \in G\}}},
 \]
 which is a infinite dimensional symplectic manifold carrying a Hamiltonian
 action of $(L G)^n$.
 The corresponding moment map is  given by the restriction map to the boundary
 components
 $\hat{\mu}_\Sigma: \hat{\cM}_\Sigma \longrightarrow (L\g^*)^n$
 defining the following commutative diagram analogous to (\ref{diagram:1}):
 \ba\label{diagram:2}
 \xymatrix{ \hat{\cM}_\Sigma\ar[d]_{\pi} \ar[r]^{\hat{\mu}_\Sigma} & (L\g^*)^n \ar[d]^{Hol}\\
     \cM_\Sigma \ar[r]^{\mu_\Sigma} &G^n.
     }
  \na
  Therefore, the pull-back bundle gerbe
 $$\mu_\Sigma^* (\otimes_{i=1}^n p_i^*\cG_k)$$
 over  $\cM_\Sigma$, where $p_i: G^n \to G$ is the projection on its $i$-th factor,
 has a canonical  trivialisation given by the pre-quantization line bundle $\cL_{\Sigma}$
 over $\hat{\cM}_\Sigma$ by a construction analogous to that in the proof of
 Proposition \ref{moduli:gb-module}. Hence,  $(\cM_\Sigma, \mu_\Sigma)$
 gives rise to a generalized equivariant bundle gerbe $D$-brane in $G^n$.
\end{proof}

\begin{remark}To see precisely the relationship between the Segal-Witten reciprocity law
and our Proposition \ref{reciprocity}, we remind the
reader of the following two observations.
\begin{enumerate}
\item Proposition \ref{reciprocity} holds for all Sobolev classes $L^2_p$
and the corresponding moduli space $\hat{\cM}_\Sigma$ contains
a dense set
$C^\infty(\partial\Sigma, G_\CC)/Hol(\Sigma, G_\CC)$
which is a Frechet manifold.
\item The pre-quantization line bundle   $\cL_{\Sigma}$ is given by
the extension of
\[
 \bigl( \widehat{C^\infty(\partial\Sigma, G_\CC)} /Hol(\Sigma, G_\CC)\bigr)\times_{\CC^*} \CC
 \]
 where the canonical splitting over $Hol(\Sigma, G_\CC)$ enters naturally.
\end{enumerate}
It is not hard   to see that these canonical  trivialisations
   obtained from the transgression
 of our universal Chern-Simons bundle  2-gerbe $\cQ_\phi$ satisfy natural
glueing properties under cutting and pasting of Riemann surfaces.
\end{remark}

\subsection{The multiplicative structure of $\cG_k$ }

We assume that the classifying map for
  the principal $G$-bundle over $\Sigma_{0, n}$ is given by a smooth map
 $ \Sigma_{0, k} \to BG $
 such that base points  on the boundary components are mapped to
 a base point in $BG$. Then the based moduli space of flat $G$-connections
 on $\Sigma_{0, n}$, still denoted by $\cM_{\Sigma_{0, n} }$,
  satisfies
 \[
 \cM_{\Sigma_{0, n} }\cong \{(g_1, \cdots, g_n) \in G^n | \prod_{i=1}^n g_i =1\}.
 \]
 For  a sphere
 with three holes $\Sigma_{0, 3}$, the pull-back bundle gerbe over
 $\cM_{\Sigma_{0, 3} }$ is isomorphic to
  the bundle gerbe
 \[
 \delta(\cG_k) =  p_0^*(\cG_k)\otimes  p_1^*(\cG^*_k)\otimes  p_2^*(\cG_k)
  \]
  over $G \times G$, where $\pi_i: G\times G \to G$
  is given by
  $\pi_0(g_1, g_2) = g_2$, $\pi_1(g_1, g_2) = g_1g_2$ and $\pi_2(g_1, g_2) = g_1$ for
  $(g_1, g_2) \in G$. Then the induced canonical trivialisation of $\delta(\cG_k)$
  defines  the multiplicative structure on $\cG_k$ (see \cite{CJMSW}).
 The associator  for the multiplicative structure  is given
 by the canonical trivialisation
 of the pull-back bundle gerbe over the based moduli space
 \[
 \cM_{\Sigma_{0, 4} }\cong G \times G \times G
 \]
 of flat $G$-connections on a
 sphere with four holes, $\Sigma_{0, 4}$,
  and the induced trivialisations from two ways of decomposing
  the four holed sphere into three holed spheres.
 The cocycle condition for the associator is given by the canonical trivialisation
 of the pull-back bundle gerbe over the based moduli space
 \[
 \cM_{\Sigma_{0, 5} }\cong G \times G \times G\times G
 \]
 of flat $G$-connections on a
 sphere with five holes, $\Sigma_{0, 5}$,
  and  various ways of decomposing  the 5-holed sphere.

   \section{$Spin^c$ quantization and fusion of $D$-branes}

Fix  a quasi-Hamiltonian $G$-manifold $(M, \omega, \mu)$
 with corresponding pre-quantizable Hamiltonian $LG$-manifold
 $(\hat{M}, \hat{\omega}, \hat{\mu})$ at level $k$, as illustrated in the following
 diagram
 \ba
 \label{LG:G}
     \xymatrix{
     (\hat{M}, \hat{\omega}) \ar[d]_{\pi} \ar[r]^{\hat{\mu}} & L\g^* \ar[d]^{Hol}\\
     (M, \omega)  \ar[r]^{\mu} &(G, \Theta_k).
     }
     \na
       From Theorem \ref{bg:module:rank1}, we know that $\hat{M}$, together with
       the pre-quantization line bundle $\cL= \cL_{\hat{M}}$, defines a generalized rank one $G$-equivariant
        bundle  gerbe module of $\cG_k$ over $G$. Equivalently, $(M, \omega, \mu)$
       is a   generalized $G$-equivariant  bundle gerbe $D$-brane of $\cG_k$. In this section, we
       will define a quantization procedure which gives rise to an element
       in $R_{k}(LG)$ for any  pre-quantizable Hamiltonian $LG$-manifold
 $(\hat{M}, \hat{\omega}, \hat{\mu})$ at level $k$.  We will apply the fusion product
 defined  in \cite{MeiWoo1} for pre-quantizable Hamiltonian $LG$-manifolds at level $k$
 to show that our quantization functor from the category of generalized $G$-equivariant
        bundle  gerbe modules of $\cG_k$ to $R_{k}(LG)$ commutes with these fusion products.

      Suppose that $\lambda $ is a quasi-regular value of $\hat{\mu}$. The Hamiltonian
 reduction at the  dominant weight  $\lambda$ of level $k$, given by
 \[
M_\lambda:=  \hat{\mu}^{-1} (\lambda) /(LG)_\lambda \cong \hat{\mu}^{-1} (\cO_\lambda)/LG,
 \]
 is a symplectic orbifold with the reduced symplectic form
 $\omega_\lambda$ and a pre-quantization line bundle with a  connection $\nabla_\lambda$
 \[
 \cL_\lambda := \cL|_{\hat{\mu}^{-1} (\lambda )} \times_{\widehat{(LG)}_\lambda}
 \CC_{(*\lambda, 1)},
 \]
 where $*\lambda$ is the dominant weight of the irreducible representation of $G$
  dual to the one with weight $\lambda$,
    $\widehat{(LG)}_\lambda$ acts on $\CC_{(*\lambda,1)}$ with weight $(*\lambda, 1)$
 (Cf. \cite{Ati} and \cite{Wood}).
 Choose an almost complex structure $J$, compatible with $\omega_\lambda$,
 which defines a canonical $Spin^c$ structure $S:=S^{\pm}_J$. Twisted by the pre-quantization line bundle
 $\cL_\lambda$, a Hermitian connection on
 $TM_\lambda$ defines a $Spin^c$ Dirac operator
 \[
 \dirac_{\lambda}: \Gamma_{L^2_k}(S^{+}\otimes \cL_\lambda) \longrightarrow
 \Gamma_{L^2_{k-1}}(S^{-}\otimes \cL_\lambda).
 \]
   The index of  $\dirac_{\lambda}$, denoted by $Index (\dirac_{\lambda}, M_\lambda)$,
 is independent of the choice of the almost complex structure and the
 Hermitian connection.
 The symplectic invariant   defined as
the index of $\dirac_{\lambda}$,
$Ind (\dirac_{\lambda})$,  is  a rational number in general
(an integer if $\hat{M}_\lambda$ is a  smooth symplectic manifold).

 \begin{remark}\begin{enumerate}
 \item If $M_\lambda$ is K\"ahler and $(L_\lambda, \nabla_\lambda )$
  is holomorphic,  then the canonical $Spin^c$ bundle and the $Spin^c$ Dirac operator
  are given by
  \[  S^{\pm} = \Lambda^{0, even/odd}(T^*M_\lambda),\qquad
  \dirac_{\lambda} =
  \sqrt{2} (\bar{\partial}_{\nabla_\lambda} +\bar{\partial}^*_{\nabla_\lambda}).
  \]
  Hence, we have
   \[
    Index (\dirac_{\lambda}, M_\lambda) = \chi (M_\lambda, \cL_\lambda),
   \]
 the Euler characteristic for the sheaf of holomorphic sections of $\cL_\lambda$.
 \item  Let  $\hat{ \cM}_{\Sigma_{0, 3}} (*\lambda, *\mu, \nu)$ be the Hamiltonian
 reduction of $\hat{ \cM}_{\Sigma_{0, 3}}$ at
 $(*\lambda, *\mu, \nu) \in (\Lambda^*_k)^3$, then the index of the $Spin^c$ Dirac operator
 on $\hat{ \cM}_{\Sigma_{0, 3}} (*\lambda, *\mu, \nu)$, see \cite{Bea} \cite{MeiWoo}
 \[
 Index(\dirac, \hat{ \cM}_{\Sigma_{0, 3}} (*\lambda, *\mu, \nu) ) = N_{\lambda,\mu}^{\nu},\]
  the fusion coefficient determined by the Verlinde
factorization formula (\cite{Ver}). The vanishing theorems for
higher cohomology groups in \cite{Tel} imply that
$N_{\lambda,\mu}^{\nu}$ agrees with the dimension of the space of
holomorphic sections of the pre-quantization line bundle over the
reduced space $\hat{ \cM}_{\Sigma_{0, 3}} (*\lambda, *\mu, \nu)$.
 \end{enumerate}
 \end{remark}

\begin{definition} \label{quantization}
The  quantization of a pre-quantizable quasi-Hamiltonian
$G$-manifold $(M, \omega, \mu)$ at level $k$ is defined
to be
\[
\chi_{k, G} (M): = \chi_{k, G} (\hat{M}) = \sum_{\lambda\in \Lambda_k^*}
Index (\dirac_{\lambda}, M_\lambda)
\cdot \chi_{\lambda, k} \in R_k(LG)
\]
where $\chi_{\lambda, k}$ is the character  of the irreducible representation
of $LG$ with highest weight $(\lambda, k)$, and $R_k(LG)$ is the Abelian group generated
by the isomorphism classes of irreducible positive energy representations
of $LG$ at level $k$. Equipped with  the fusion product
\[
\chi_{\lambda, k} \ast \chi_{\mu, k} = \sum_{\nu\in \Lambda_k^*}
N_{\lambda,\mu}^{\nu}\chi_{\nu, k}.
\]
 $(R_k(LG), \ast)$ becomes the Verlinde ring.
\end{definition}

It was shown in \cite{FHT} that $(R_k(G), \ast)$ can be identified with
the $G$-equivariant twisted $K$-group $K^{dim G}_{G, h^\vee}(G)$
for the conjugacy
action of $G$ on itself, where $h^\vee$ is the
dual Coxeter number. This motivates  the
following definition.

\begin{definition}
The category of equivariant bundle gerbe $D$-branes of the equivariant bundle gerbe $\cG_k$
over $G$ is given by the category of
pre-quantizable quasi-Hamiltonian $G$-manifolds whose objects are $(M, \omega, \mu)$ and
the morphism between
$ (M_1, \omega_1, \mu_1)$ and $ (M_2, \omega_2, \mu_2)$ is given by
a $G$-equivariant map $f: M_1 \to M_2$ such that
\[
\omega_1 = f^* \omega_2, \qquad \mu_1 = \mu_2 \circ f.
\]
Equivalently, we say that
the category  of  equivariant bundle gerbe modules of  $\cG_k$   is
given by the category of pre-quantizable Hamiltonian $LG$-manifolds
 at level $k$ with proper moment maps.  We denote
this category by $\cQ_{G, k}$.
\end{definition}

On the category of pre-quantizable Hamiltonian $LG$-manifolds
at level $k$ with proper moment maps, there is a product structure,
called the {\bf fusion product of Hamiltonian $LG$-manifolds} in \cite{MeiWoo1}. Recall
from Section \ref{section:moduli} that $\cM_{\Sigma_{0, 3}}$ is the based
moduli space of flat $G$-connections on $\Sigma_{0, 3}$ (the genus $0$ surface
with 3 pointed boundary components).  Note that
$\cM_{\Sigma_{0, 3}}$ is a quasi-Hamiltonian $(G\times G \times G)$-manifold,
and the corresponding Hamiltonian $(LG)^3$-manifold at level $k$ is denoted by
$\hat{ \cM}_{\Sigma_{0, 3}}$:
\[
\xymatrix{ \hat{\cM}_{\Sigma_{0, 3}}\ar[d]_{\pi}
\ar[r]^{\hat{\mu}_{\Sigma_{0, 3}}} & (L\g^*)^3 \ar[d]^{Hol}\\
     \cM_{\Sigma_{0, 3}} \ar[r]^{\mu} &G^3.
     }
\]

Given a Hamiltonian $LG \times LG$-manifold $\hat M$ at level $k$,
\[
\xymatrix{\hat{\cM} \ar[d]  \ar[r]^{\hat{\mu}} &  L\g^*\oplus L\g^* \ar[d]^{Hol}\\
     \cM_{\Sigma_{0, 3}} \ar[r]^{\mu} &G\times G }
\]
then $\hat{M}\times \hat{M}_{\Sigma_{0,3}} $ is a Hamiltonian
$(LG)^5$-manifold at level $k$. The diagonal embedding
\[
LG \times LG \longrightarrow (LG \times LG)  \times (LG \times LG
\times LG),
\]
mapping $(\gamma_1, \gamma_2) \mapsto (\gamma_1, \gamma_2) \times
(\bar{\gamma}_1, \bar{\gamma}_2, e)$ (where $\bar{\gamma}_i
(\theta) = \gamma_i (-\theta): \RR/\ZZ \to G$), defines a
Hamiltonian $LG \times LG$ action on $\hat{M}\times
\hat{M}_{\Sigma_{0,3}} $ with moment map
\[\begin{array}{cccc}
\hat{\mu}_{diag}: & \hat{M}\times \hat{M}_{\Sigma_{0,3}}&
\longrightarrow & L\g^* \oplus L\g^*\\[2mm]
& (x, [A] ) &\mapsto & \hat{\mu}(x) - pr_{12} \circ
\hat{\mu}_{\Sigma_{0, 3}}([A])
\end{array}
\]
where $pr_{12}$ denotes the projection from $L\g^* \oplus L\g^*
\oplus L\g^* $ to the first two factors. As $0$ is a regular value
of  $pr_{12} \circ \hat{\mu}_{\Sigma_{0, 3}}$, we can define the
Hamiltonian quotient, denoted by
\[
\hat{M}\times \hat{M}_{\Sigma_{0,3}}// diag( LG)^2,
\]
as the symplectic   reduction
\[
\hat{\mu}_{diag} ^{-1}(0) / (LG \times LG) .
\]
Note that $$\hat{\mu}_{diag}^{-1}(0) / (LG \times LG) \cong
\bigl(\hat{M}\times_{L\g^*\oplus L\g^*}
\hat{M}_{\Sigma_{0,3}}\bigr)/(LG \times LG), $$ which is an
 $LG \times LG$ quotient
of  the fiber product of
$\hat{\mu}: \hat{M} \to L\g^* \oplus L\g^*$ and $pr_{12} \circ
\hat{\mu}_{\Sigma_{0, 3}}: \hat{M}_{\Sigma_{0,3}} \to
 L\g^* \oplus L\g^*$.
The remaining $LG$-action on $\hat{M}_{\Sigma_{0,3}}$ descends to
a Hamiltonian $LG$-action on $\hat{M}\times
\hat{M}_{\Sigma_{0,3}}// diag( LG)^2$,  with the natural moment
map induced from the map
\[
\begin{array}{cccc}
pr_3 \circ \hat{\mu}_{\Sigma_{0, 3}}& \hat{M}\times
\hat{M}_{\Sigma_{0,3}}&
\longrightarrow & L\g^* \oplus L\g^*\\[2mm]
 &(x, [A] ) &\mapsto &  pr_3 \circ
\hat{\mu}_{\Sigma_{0, 3}}([A])
\end{array}
\]
where $pr_3$ is the projection from $L\g^* \oplus L\g^* \oplus
L\g^*$ to the third factor. As any $\lambda \in L\g^*$ is a
regular value of $pr_3 \circ \hat{\mu}_{\Sigma_{0, 3}}$, we can
define the symplectic reduction of $\hat{M}\times
\hat{M}_{\Sigma_{0,3}}// diag( LG)^2$ at $\lambda$ as
\[
(pr_3 \circ \hat{\mu}_{\Sigma_{0, 3}})^{-1}(0)/(LG)_\lambda \cong
\bigl(\hat{M}\times \hat{M}_{\Sigma_{0,3}} (\cdot, \cdot,
\lambda)\bigr)/diag(LG)^2,
\]
where $\hat{M}_{\Sigma_{0,3}} (\cdot, \cdot, \lambda)$ is given by
the subset of $\hat{M}_{\Sigma_{0,3}}$ with holonomy around the
outgoing boundary component in the conjugacy class ${\cC}_\lambda$ of  $G$
 through $\exp({{2\pi i\lambda}/{k}})$.

\begin{definition}
The {\bf fusion product} on the category $\cQ_{G, k}$ is defined as follows: given
two pre-quantizable Hamiltonian $LG$-manifolds
$(\hat{M}_1, \hat{\omega}_1, \hat{\mu}_1)$  and $(\hat{M}_2, \hat{\omega}_2, \hat{\mu}_2)$ at
level $k$
with proper moment maps, the fusion of product
$\hat{M}_1 \boxtimes  \hat{M}_2$ is the Hamiltonian $LG$-manifold at level $k$
 obtained as the Hamiltonian quotient
\[
\hat{M}_1 \boxtimes \hat{M}_2:= \bigl( (\hat{M}_1 \times \hat{M}_2) \times
\hat{ \cM}_{\Sigma_{0, 3}} \bigr) //diag(LG)^2,
\]
with the resulting moment map denoted by $\hat{\mu}_1 \star \hat{\mu}_2$.
For two pre-quantizable quasi-Hamiltonian $G$-manifolds $(M_1, \omega_1, \mu_1)$ and
$(M_2, \omega_2, \mu_2)$, the corresponding fusion product
is given by
\[
M_1\boxtimes M_2 = (M_1\times M_2, \omega_1 + \omega_2
+ \disp{\frac k2} <\mu^*_1\theta, \mu_2^*\bar{\theta}>, \mu_1\cdot \mu_2).\]
\end{definition}

\begin{remark}
By direct calculation, it is shown in \cite{MeiWoo1} that
\[
\hat{ \cM}_{\Sigma_{g_1, n_1}} \boxtimes \hat{ \cM}_{\Sigma_{g_2, n_2}}
= \hat{ \cM}_{\Sigma_{g_1+ g_2, n_1+ n_2 -1}},
\]
and, for three level $k$  dominant weights $\lambda$, $\mu$ and $\nu$ in
$\Lambda^*_k$,
\[
(\cO_\lambda \boxtimes \cO_\mu )_\nu = \hat{\cM}_{\Sigma_{0, 3}}(*\lambda, *\mu, \nu),
\]
where $\cO_\lambda$ denotes the coadjoint orbit of the affine $LG$
action on $L\g^*$ at level $k$ with the corresponding
quasi-Hamiltonian $G$-manifold given by
\[
\cC_\lambda = \{ g\cdot exp(\disp{\frac{2\pi i\lambda}{k}})\cdot
g^{-1}| g \in G\}.
\]
\end{remark}

The fusion product on  $\cQ_{G, k}$ is well-defined in the sense that  given two
pre-quantizable quasi-Hamiltonian $G$-manifolds $(M_1, \omega_1, \mu_1)$ and
$(M_2, \omega_2, \mu_2)$ whose Hamiltonian $LG$-manifolds at level $k$ are denoted
by $\hat{M}_1$ and  $\hat{M}_2$, then
the corresponding Hamiltonian $LG$-manifold at level $k$  for $M_1\boxtimes M_2$ is
given by $$\widehat{M_1\boxtimes M_2} = \hat{M}_1 \boxtimes \hat{M}_2,$$
which is also pre-quantizable. See \cite{AAM} for a proof of this claim.
 Moreover, modulo $LG$-equivariant
symplectomorphisms, $\cQ_{G, k}$ is a monoidal tensor category:
\begin{enumerate}
\item For any Hamiltonian $LG$-manifold $\hat{M}$ at level $k$,  there is an $LG$-equivariant
symplectomorphism $\hat{M}\boxtimes \Omega G \cong \hat{M}$, that is,
$\Omega G$ is the unit object.
\item Let $\hat{M}_1, \hat{M}_2, \hat{M}_3$ be Hamiltonian $LG$-manifolds at level $k$
with proper moment maps. There exist $LG$-equivariant
symplectomorphisms
\[ \hat{M}_1\boxtimes \hat{M}_2 \cong \hat{M}_2\boxtimes \hat{M}_1,
\]
\[
(\hat{M}_1\boxtimes \hat{M}_2)\boxtimes \hat{M}_3 \cong \hat{M}_1\boxtimes
 (\hat{M}_2\boxtimes \hat{M}_3).
\]
\end{enumerate}

The category  $\cQ_{G, k}$ together with the fusion product
is called the fusion category of equivariant bundle gerbe $D$-branes $(\cQ_{G, k}, \boxtimes)$.
Our main theorem about the structure of
the category $\cQ_{G, k}$ is the following result on
``quantization commutes with fusion",  which
gives a geometric way to think of the ring structure on
equivariant twisted $K$-theory (cf  \cite{FHT}).

\begin{theorem}\label{main:theorem}
The quantization  functor defined by the $Spin^c$ quantization as
in Definition \ref{quantization}
$\chi_{k, G}: (\cQ_{G, k}, \boxtimes)  \longrightarrow (R_k(LG), \ast)$
satisfies
\ba\label{fusion:commute}
\chi_{k, G} (\hat{M}_1\boxtimes \hat{M}_2) = \chi_{k, G} (\hat{M}_1 )\ast\chi_{k, G} (\hat{M}_2 ),
\na
where the product $\ast$ on the right hand side denotes the fusion ring structure
on the Verlinde ring $R_k(G)$.
\end{theorem}
\begin{proof} By definition, we see that
\[\chi_{k, G} (\hat{M}_1\boxtimes \hat{M}_2)
= \sum_{\nu\in \Lambda_k^*} Index \bigl( \dirac_\nu, (\hat{M}_1\boxtimes \hat{M}_2)_\nu \bigr)
\cdot \chi_{\nu, k},
\]
and
\[\chi_{k, G} (\hat{M}_1 )\ast\chi_{k, G} (\hat{M}_2 )
= \sum_{\lambda, \mu, \nu \in \Lambda_k^*}  Index
\bigl(\dirac_\lambda, (\hat{M}_1)_\lambda\bigr) \cdot Index
\bigl(\dirac_\mu, (\hat{M}_2)_\mu\bigr)\cdot
N_{\lambda,\mu}^\nu\cdot\chi_{\nu, k}.
\]

Note that for $\nu\in \Lambda_k^*$
\[
(\hat{M}_1\boxtimes \hat{M}_2)_\nu
= \bigl( (\hat{M}_1 \times \hat{M}_2) \times
\hat{ \cM}_{\Sigma_{0, 3}}(\nu)  \bigr) //diag(LG)^2.\]
Applying Theorem 2.1 in \cite{MeiWoo} to the
Hamiltonian $L(G\times G \times G \times G)$-manifold at level $k$
\[
(\hat{M}_1 \times \hat{M}_2) \times \hat{ \cM}_{\Sigma_{0, 3}}(\nu),
\]
we obtain
\[\begin{array}{lll}
&&Index \bigl( \dirac_\nu, (\hat{M}_1\boxtimes \hat{M}_2)_\nu \bigr) \\[2mm]
&=&\sum_{\lambda, \mu \in \Lambda_k^*} Index \bigl(\dirac,
( (\hat{M}_1 \times \hat{M}_2) \times \hat{ \cM}_{\Sigma_{0, 3}}(\nu))_{\lambda, \mu,
*\lambda, *\mu} \bigr)
\\[2mm]
&=&\sum_{\lambda, \mu \in \Lambda_k^*} Index \bigl(\dirac,
(\hat{M}_1 \times \hat{M}_2)_{\lambda, \mu} \times \hat{ \cM}_{\Sigma_{0, 3}}(*\lambda, *\mu, \nu)
\bigr)
\\[2mm]
&=& \sum_{\lambda, \mu \in \Lambda_k^*} Index \bigl(\dirac_\lambda, (\hat{M}_1)_\lambda\bigr)
\cdot Index \bigl(\dirac_\mu, (\hat{M}_2)_\mu\bigr) \cdot
Index \bigl(\dirac, \hat{ \cM}_{\Sigma_{0, 3}} (*\lambda, *\mu, \nu) \bigr)\\[2mm]
&=& \sum_{\lambda, \mu \in \Lambda_k^*\ } N_{\lambda,\mu}^\nu\cdot
Index \bigl(\dirac_\lambda, (\hat{M}_1)_\lambda\bigr) \cdot Index
\bigl(\dirac_\mu, (\hat{M}_2)_\mu\bigr),
\end{array}\]
which leads to (\ref{fusion:commute}) by direct calculation.
\end{proof}

For a dominant weight $\lambda$ of level $k$ in $\Lambda^*_k$, the
pre-quantizable quasi-Hamiltonian $G$-manifold given by the conjugacy class
$\cC_\lambda$ defines an object in $\cQ_{k, G}$. Then it is easy
to see that
\[
\chi_{k, G} (\cC_\lambda) = \chi_{k, G} (\cO_\lambda) = \chi_{\lambda, k} \in
R_k(LG).
\]
Hence, the quantization functor $\chi_{k, G}: \cQ_{k, G} \to R_k(LG)$
is surjective.

\section{The non-simply connected case}

 For a compact, connected, non-simply connected simple Lie group $G$,
there are a few subtleties
 in the construction of the fusion category of bundle gerbe modules.
 These issues also surface in the ring structure
 on the equivariant twisted K-theory of $G$, see for example \cite{CW2}.

 Let $\tilde G$ be the universal cover of $G$,
$G= \tilde{G}/Z$, where $Z$ is a
  subgroup of the center $Z(\tilde G)$ of
 $\tilde G$,
 \[
 1\to Z \longrightarrow\tilde G \longrightarrow G \to 1.
 \]
  and the covering map $\pi: \tilde G \to G$ identifies $Z$ with
  the fundamental group $\pi_1(G)$. For a compact, connected and simply connected
  simple Lie group with non-trivial center, $\tilde G$ is one of the Cartan
  series $SU(n+1)$, $Spin(2n+1)$, $Sp(n)$,  $Spin(4n)$, $Spin(4n+2)$, $E_6$ and $E_7$
  with the  center  given by $\ZZ_{n+1}$, $\ZZ_2$, $\ZZ_2$, $\ZZ_2\times \ZZ_2$,
  $\ZZ_4$, $\ZZ_3$ and $\ZZ_2$ respectively.

The first  subtlety comes from the non-connectness
of the loop group $LG$, in fact its
connected components are given by the fundamental group $\pi_1(G)$.
Let
$L_0 G$ be the identity component of $LG$, and
$\Omega_0 G$  the identity component of $\Omega G$, then we have the following exact sequences:
\[1\to L_0G \longrightarrow LG \longrightarrow  Z \to 1,\quad\quad
1\to \Omega_0G \longrightarrow \Omega G \longrightarrow  Z \to 1.
\]
We now note the following  difficulties:
\begin{enumerate}
\item The central extension of $LG$,
$1\to  U(1) \longrightarrow \widehat{LG} \longrightarrow LG \to 1,$
is not uniquely determined by a class $\sigma \in H^3(G, \ZZ)$, rather
by a class in $H^3(G, \ZZ) \oplus Hom (Z, U(1))$ (\cite{PS} and
\cite{Tol}).
\item The fusion object for the category of bundle
gerbe modules for
a non-simply connected $G$,  as a moduli space of flat connections on $\Sigma_{0, 3}$ modulo
those gauge transformations which  are trivial on boundary components,   is actually
a  Hamiltonian $L_0 G \times L_0 G \times L_0 G$-manifold at level $k$,
 not a Hamiltonian $LG \times LG \times LG$ manifold.
 \item The pre-quantization condition for Hamiltonian $L_0 G$-manifolds at level $k$, such as those
 from the  moduli spaces of flat connections on a Riemann surface,
only holds when the level $k$ is transgressive for $G$.
 \end{enumerate}

The second subtlety comes from the fact that we have to
 restrict to bundle gerbes $P_\sigma$ whose
Dixmier-Douady class $\sigma$ lies in
the image of $\tau: H^4(BG, \ZZ) \to H^3(G, \ZZ)$ in order that
$P_\sigma$ is multiplicative.

\begin{remark}
\label{level}\begin{enumerate}
\item For $G= SO(3)$, a Dixmier-Douady class $\sigma$
is transgressive if it is an  even class in
$H^3(SO(3), \ZZ)$, equivalently,
a multiple of $4$ under the map
$$H^3(SO(3), \ZZ) \longrightarrow H^3(SU(2), \ZZ) \cong \ZZ.$$
\item For a general compact, connected, simple Lie group $G$, we have
 the following commutative diagram:
\ba\label{trans}
\xymatrix{ H^4(BG, \ZZ) \ar[d]  \ar[r]  & H^3(G, \ZZ) \ar[d] \\
   H^4(B\tilde G, \ZZ)   \ar[r]^{\cong} & H^3(\tilde G, \ZZ),
     }
     \na
 where $  H^4(B\tilde G, \ZZ) \cong  H^3(\tilde G, \ZZ) \cong \ZZ$
 and the generator is determined by the basic inner product
$<\cdot, \cdot>$   on $\g$ uniquely specified
by requiring that the highest co-root of  $\tilde G$  has norm $2$.
Then each element in
$H^4(B\tilde G, \ZZ)$
is specified by its ``level'' $k$
coming from identifying the induced  inner product on $\g$ as being
given by $k <\cdot, \cdot>$.
\end{enumerate}
\end{remark}

\subsection{Some background about the center of $\tilde G$ and the loop group $LG$}

We first review some basic properties about the center of
the universal cover $\tilde G$ of $G$ following  \cite{Tol} where
irreducible positive energy projective representations of $LG$ are
classified.

  The center $Z(\tilde G)$  can be characterized as follows:
choose a maximal torus $\tilde T$ of $\tilde G$ with Lie algebra
$\t$, let $R$ and $R^\vee$ be the root system and the co-root
system of $\tilde G$. The root and co-root lattices $\Lambda_r
\subset  \t^*$ and $\Lambda_r^\vee\subset  \t $ are the lattices
spanned by
  $R$ and $R^\vee$ with
their $\ZZ$-basis given by
\[
\Delta = \{ \a_1, \cdots, \a_n\}, \ \ and \ \ \Delta^\vee= \{
\a_1^\vee, \cdots, \a_n^\vee\}\] respectively, where $\a^\vee  =
\disp{\frac{2 \a}{<\a, \a>}}$.   The weight and co-weight lattices
$\Lambda_w \subset  \t^*$ and $\Lambda_w^\vee\subset  \t $ are the
lattices dual to $\Lambda_r^\vee$ and  $\Lambda_r$ with their
$\ZZ$-basis given by the fundamental weights $\lambda_i$ and the
fundamental co-weights $\lambda_i^\vee$ such that
\[
<\lambda_i, \a_j^\vee > = <\lambda_i^\vee, \a_j  > = \delta_{ij}.
\]
Under the identification $\t\cong \t^*$  defined by the basic
inner product,
\[
\begin{array}{ccccc}
\Lambda_r & \subset & \Lambda_w & \subset & \t^*\\
\cup & & \cup &&  \\
\Lambda_r^\vee & \subset & \Lambda_w^\vee &\subset & \t,
\end{array}
\]
from which we know that
\begin{enumerate}
\item The map $h^\vee \mapsto exp_{\tilde T}(2\pi i h^\vee)$
induces an isomorphism $ \Lambda_w^\vee/ \Lambda_r^\vee \cong
Z(\tilde G)$, where $exp_{\tilde T}$ denotes the exponential map
for $\tilde T$.
 \item The map sends $\mu\in\Lambda_w $ to the  pairing
$\mu(exp_{\tilde T}(2\pi h^\vee)) = e^{2\pi i <\mu, h^\vee>}$ induces an
isomorphism $ \Lambda_w/ \Lambda_r  \cong Hom (Z(\tilde G), U(1))$.
\end{enumerate}
There is another characterisation of $Z(\tilde G)$ in terms of
special roots (Cf. Lemma 2.3 in \cite{Tol}): non-trivial elements
in $Z(\tilde G)$ correspond one-to-one  to the special
fundamental co-weights
$$\{\lambda_{i(z)}\}_{z \in Z(\tilde G)}$$
such that the corresponding $\a_{i(z)}\in \Delta$ carries the
coefficient $1$ in the expression for the highest root $\vartheta$.
For each special root $\a_{i(z)}$, there exists a unique Weyl
group element $\omega_{i(z)}$ (Cf. Proposition 4.1.2 in
\cite{Tol}) which preserves $\Delta\cup \{-\vartheta\}$ and sends
$-\vartheta$ to $\a_{i(z)}$, such that
\[
\omega_{i(z_1)}\omega_{i(z_2 )} = \omega_{i(z_1 \cdot
z_2)},
 \]
 where $z_1 \cdot z_2$ denotes the group multiplication
 in $Z(\tilde G)$.

 The dominant weight of $\tilde G$ at level $k$ is given by
 \ba\label{level-k:weights}
\Lambda_k^* = \{\lambda \in \Lambda_w | <\lambda, \alpha^\vee>
\geq 0, <\lambda, \theta> \leq k\} \na which is non-empty only if
$k$ is a positive integer.
 Note that $Z(\tilde G)$ is
isomorphic to the group of automorphisms of the extended Dynkin
diagram of $\tilde G$ which induces an action of $Z(\tilde G)$
on $\Lambda_k^*$ as given by Proposition 4.1.4 in \cite{Tol},
where the explicit action for all classical groups is explained.
Geometrically, this action of
$Z(\tilde G)$ is given by the  affine Weyl group element (\cite{Tol})
\ba\label{action:affine}
z \mapsto  \tau(k\lambda^\vee_{i(z)})\omega_{i(z)},
\na
where $\tau (k\lambda^\vee_{i(z)})$ denotes the translation by
$k\lambda^\vee_{i(z)}$ in the affine Weyl group.

Given a subgroup $Z\subset Z(\tilde G)$,  the integral lattice of
$G = \tilde G /Z$,
  $\Lambda_Z^\vee = Hom (U(1), \tilde T/Z)$, then
 $\Lambda_r^\vee \subset \Lambda_Z^\vee \subset \Lambda_w^\vee$, and
 $\Lambda_Z^\vee / \Lambda_r^\vee \cong Z$. The basic level $\ell_b$
 of $G$ is the smallest integer $k$ such that the restriction
 of $k <\cdot, \cdot >$ to $\Lambda_Z^\vee$ is integral.

Introduce the group of `discontinuous loops'
 \[
 L_Z \tilde G = \{ \gamma\in C^\infty (\RR, \tilde G) | \gamma(t +2\pi)
 \gamma (t)^{-1} \in Z\}.
 \]
 Then we have the following commutative diagram
 with all rows and columns being exact:
 \ba\label{row:col}
 \xymatrix{
 & 1\ar[d] & 1\ar[d] & & \\
 & Z \ar[r]^{\cong}\ar[d] & Z \ar[d]& & \\
 1\ar[r] & L\tilde G \ar[d]\ar[r] &L_Z\tilde G \ar[d]\ar[r] & Z \ar[r]\ar[d]^{\cong}
  & 1\\
 1\ar[r]   & L_0 G \ar[d]\ar[r] &LG \ar[d]\ar[r] & Z \ar[r] &1\\
 & 1 & 1 & &}
 \na

 As $Z\cong \Lambda_Z^\vee/\Lambda_r^\vee$, we can associate to
$\mu^\vee \in \Lambda_Z^\vee$, the   discontinuous
loop
\ba\label{dis}
\zeta_{\mu^\vee}(t) = exp_{\tilde T}(2\pi i t \mu^\vee)
\in L_Z(\tilde G).
\na
Notice that if $\mu^\vee \in \Lambda_r^\vee$, then
$\zeta_{\mu^\vee}(t) \in L\tilde T$. 
In particular, for each $z\in Z$, fix a representative
$w_z\in \tilde{G}$ for the  unique Weyl
group element $\omega_{i(z)}$, then 
\ba\label{distinguished}
z\mapsto \zeta_z: = \zeta_{\lambda^\vee_{ i(z)}}w_z 
\na
assigns a discontinuous
loop in  $L_Z(\tilde G)$ to each $z$, we call $\zeta_z$ the distinguished discontinuous loop
associated to $z$.  According to Theorem 4.3.3 in \cite{Tol}, the conjugation action of
$\zeta_z$  on $L\tilde G$  induces an action of $Z$ on
$\Lambda_k^*$ which agrees with the induced action of $Z \subset
Z(\tilde G)$ on $\Lambda_k^*$ as from (\ref{action:affine}).  We
denote by $z\cdot \lambda$ the action of $z \in Z$ on
$\lambda \in \Lambda_k^*$.

\begin{definition} The basic level  $\ell_b$ of $G=\tilde{G}/Z$ is the smallest
positive integer $\ell$ such that the restriction of $\ell<\cdot, \cdot>$ to
$\Lambda_Z^\vee$ is integral.
\end{definition}

It was shown in Proposition 3.5.1 of \cite{Tol}, the level for which
$LG$ admits a central extension is a multiple of  the basic level $\ell_b$.
Now we review the construction of the central extension of $LG$ from \cite{Tol}.

 Given a level $k \in \ell_b \ZZ$, there is a canonical central extension
 $\widehat{L_Z \tilde G}$ of $L_Z \tilde G$  (see
 Proposition 3.5.1 and Theorem 3.2.1 in \cite{Tol}) at level $k$
 and the trivial extension of $\Lambda^\vee_Z$
 where we regard $\Lambda^\vee_Z$  as a subgroup of $L_Z \tilde G$  through
 the   discontinuous loop $\zeta_{\mu^\vee}$ given by
 (\ref{dis}) for $ \mu^\vee \in\Lambda^\vee_Z$.
 The  central extension of $\Lambda^\vee_Z$ is classified by its commutator map
 defined by
 \ba\label{commutator}
 \omega (\lambda^\vee, \mu^\vee) =
 \tilde{\zeta}_{\lambda^\vee} \tilde{\zeta}_{\mu^\vee}
 \tilde{\zeta}_{\lambda^\vee}^{-1}\tilde{\zeta}_{\mu^\vee}^{-1}
 \na
 where $\tilde{\zeta}_{\lambda^\vee},\tilde{\zeta}_{\mu^\vee} \in \widehat{L_Z \tilde G}$
 are arbitrary lifts of $\tilde{\zeta}_{\lambda^\vee}, \tilde{\zeta}_{\mu^\vee}$.
 Note that there is a  necessary and sufficient compatibility condition
 (Cf. Proposition 3.3.1)
 \[
 \omega (\lambda^\vee, \mu^\vee) = (-1)^{k<\lambda^\vee, \mu^\vee>}
 \]
 for the existence of $\widehat{L_Z \tilde G}$, whenever $\lambda^\vee \in \Lambda_r^\vee$.

 As $\tilde G$ is simple and  simply-connected, the restriction
 of $\widehat{L_Z \tilde G}$ to $\tilde G$, is canonically trivial, hence,
 restricted to $Z$, $\widehat{Z}$ admits a canonical
 section $s: Z \to \widehat{Z}$. Given a character $\chi: Z \to U(1)$,
 following \cite{Tol}, we can construct a canonical
   central extension of $LG$ associated to a level $k \in \ell_b\ZZ$ and
 $\chi   \in Hom (Z, U(1))$, defined by
 \ba\label{LG:extension}
 \widehat{LG}_\chi : = \widehat{L_Z \tilde G}/\{s(\gamma) \chi^{-1}(\gamma)| \gamma \in Z\}.
 \na
We denote this central extension of $LG$ by
\ba\label{LG:alpha}
1\to  U(1) \longrightarrow \widehat{LG}_\chi \longrightarrow LG \to 1.
 \na

 \begin{remark}\label{PSO}
  For all compact, connected simple Lie groups $G$ except
 $$
 PSO(4n) = \disp{\frac {Spin(4n)}{\ZZ_2 \times \ZZ_2}},
 $$
 we have $H^3(G, \ZZ)\cong \ZZ$. Then from (\ref{LG:extension}) we obtain all
 central extensions of $LG$ labelled by $(k, \chi)$. For $G = Spin(4n)/Z$
 with $Z=\ZZ_2 \times \ZZ_2$,
 $$ H^3(G, \ZZ)\cong \ZZ \oplus \ZZ_2,$$
 where $\ZZ_2$ corresponds to two inequivalent central extensions of $\widehat{L_Z \tilde G}$:
 the canonical one for $\widehat{LG}_\chi$ and the other one with the
 commutator $\omega$ (\ref{commutator})  defined by  the pull-back of the non-trivial,
 skew-symmetric form on $\ZZ_2 \times \ZZ_2$.
 We denote the corresponding central extension of $LG$ with respect  to this
 non-trivial commutator $\omega$ by
 \[
 1\to  U(1) \longrightarrow \widehat{LG}_{\chi, -} \longrightarrow LG \to 1.
 \]
The following discussion for $\widehat{LG}_{\chi}$
can be extended to the case of $\widehat{LG}_{\chi, -}$
 for $G = PSO(4n)$ with some minor modifications, we shall point out the difference for this latter case.
 \end{remark}

 We are now able to classify all   irreducible positive energy
 representations of $ \widehat{LG}_\chi$.  We assume from now on that
the level $k \in \ell_b\ZZ$.
 The induced central  extensions of
 $L_0G$ and $L\tilde G$ from $\widehat{LG}_\chi$ and $\widehat{L_Z \tilde G}$
 are denoted by $\widehat{L_0G}_\chi$ and $\widehat{L\tilde G}$ respectively.
  Then we have the following commutative diagrams
relating various exact sequences and their central extensions:
\ba\label{LG:L0G}
\xymatrix{
1\ar[r]& \widehat{L_0G}_\chi \ar[r]\ar[d]& \widehat{LG}_\chi \ar[r]\ar[d]& Z\ar[d]^{\cong}
 \ar[r] & 1 \\
 1\ar[r]&L_0G \ar[r] &LG\ar[r]  & Z \ar[r] & 1,
}
\na
\ba\label{L1G:L0G}
\xymatrix{
1\ar[r]& Z   \ar[r]\ar[d] & \widehat{L\tilde G} \ar[r]\ar[d]&
 \widehat{L_0G}_\chi \ar[d]
 \ar[r] & 1 \\
 1\ar[r]&Z   \ar[r] &L\tilde G \ar[r]  &  L_0G \ \ar[r] & 1,
}
\na
 and
 \ba\label{LG:LZG}
\xymatrix{
1\ar[r]& Z   \ar[r]\ar[d] & \widehat{L_Z\tilde G} \ar[r]\ar[d]&
 \widehat{LG}_\chi \ar[d]
 \ar[r] & 1 \\
 1\ar[r]&Z   \ar[r] &L_Z\tilde G \ar[r]  &  L G \ \ar[r] & 1.
}
\na

 The key observation is the following proposition which characterizes
  irreducible positive energy representation of $
  \widehat{LG}_\chi$ and Theorem 4.3.3 of \cite{Tol}. The proof of
  Proposition \ref{rep:chi} follows directly from
  the definition of $ \widehat{LG}_\chi$ in (\ref{LG:extension}).

  \begin{proposition}\label{rep:chi}
   For  any irreducible positive energy
 representation of $ \widehat{LG}_\chi$, the pull-back representation to
 $\widehat{L_Z\tilde G}$ is an irreducible positive energy
 representation of $\widehat{L_Z\tilde G}$
(as classified  in \cite{Tol})
 such that the center $Z$, as a subset  of $ \widehat{L_Z\tilde G}$ under the
 canonical section $s: Z \to \hat{Z}$, acts as multiplication by the
 character $\chi$.
 \end{proposition}

\begin{proposition}\label{4.3.3}
(Theorem 4.3.3 of \cite{Tol})
 Let $(\cH_\lambda, \pi)$ be an
irreducible positive energy representation of $\widehat{L\tilde
G}$ of level $k$ and highest weight $\lambda \in \Lambda_k^*$.
For the distinguished discontinuous loop $\zeta_z$  associated to $z  \in  Z $, 
 the conjugated representation
$$\gamma \mapsto \pi( \zeta_z^{-1} \gamma  \zeta_z)$$
of $\widehat{L\tilde G}$ on $H_\lambda$ is an irreducible positive
energy representation of $\widehat{L\tilde G}$ of level $k$ and
the highest weight $z\cdot \lambda \in \Lambda_k^*$,
where $ z\cdot \lambda$ denotes  $Z\subset Z(\tilde G)$-action as in (\ref{action:affine}).
 \end{proposition}

Given  a distinguished discontinuous
loop $\zeta_z \in L_Z\tilde G$ associated to $z \in Z$,  from Proposition
\ref{4.3.3}, we know that the conjugated representation of
$(\cH_\lambda, \pi)$ by $\zeta_z$, denoted by $(\zeta_z)_*
\cH_\lambda$, is equivalent to the irreducible positive energy
module $\cH_{z \cdot \lambda}$ of $\widehat{L\tilde
G}$.

There is a character map $\Lambda_k^* \longrightarrow Hom (Z,
U(1))$ given by \ba \label{character} \lambda \mapsto e^{2\pi
i<\lambda, \cdot >}, \na where $e^{2\pi i<\lambda, \cdot >}$ is
the character on $Z$: $h^\vee \mapsto  e^{2\pi i <\lambda,
h^\vee>} $ for $h^\vee \in \Lambda_Z^\vee$.

\begin{lemma} (Cf. Lemma 7.1 in \cite{Tol}) For any positive
energy irreducible representation $\cH_\lambda$ of
$\widehat{L\tilde {G}}$ at level $k$, the unique lift of $Z\subset
\tilde {G}$ acts on $\cH_\lambda$ by the character
\[
h^\vee \mapsto e^{2\pi i < \lambda, h^\vee >}
\]
for $[h^\vee ] \in \Lambda_Z^\vee/\Lambda_r^\vee \cong Z$.
\end{lemma}

Denote  by $\Lambda_{k, \chi}^*$ the pre-image of $\chi \in
Hom(Z, U(1))$ for the  character map (\ref{character}). Then
$\Lambda_k^*$ is partitioned into different sectors labelled by
$\chi \in  Hom(Z, U(1))$:
\[
\Lambda_k^*  = \bigcup_{\chi \in  Hom(Z, U(1))} \Lambda_{k,
\chi}^*.
\]
Note that the character map (\ref{character}) factors through
$\Lambda_k^*/Z$, the orbit space of the  $Z$-action on
$\Lambda_k^*$ (this follows from  Lemma 7.2 and the proof of
Corollary 7.3 in \cite{Tol}).  We fix a choice of a representative
$\lambda$   for each $Z$-orbit $Z\cdot \lambda $.

The irreducible positive energy  representations of $\widehat{LG}_\chi$ are 
labelled by elements in  the space of orbits for $\Lambda_{k, \chi}^*/Z$ (Cf. Theorem
6.1 and Corollary 7.3 in \cite{Tol}), together with 
 a character $\rho \in Hom(Z_\lambda, U(1))$  for an orbit $Z\cdot
 \lambda$ with a non-trivial
  stabilizer
 \[
 Z_\lambda = \{ z \in Z| z \cdot \lambda = \lambda\}.
 \]

  Given an  orbit
 $$Z\cdot \lambda \in \Lambda_{k, \chi}^*/Z,$$
 and $\rho \in Hom (Z_\lambda, U(1))$,  denote by $\cH^\rho_{Z\cdot \lambda}$
 the  irreducible positive energy
 module of $\widehat{LG}_\chi$.  The pull-back representation, as
 a $\widehat{L\tilde G}$-module,  through the compositions of maps
   \[\xymatrix{
   \widehat{L\tilde G} \ar[r] \ar[d] &\widehat{L_Z\tilde G} \ar[d]\\
   \widehat{L_0 G}_\chi\ar[r]  &\widehat{LG}_\chi
   }
   \]
    admits the following  decomposition
 \ba
 \label{LG:decom}
\cH^\rho_{Z\cdot \lambda}\cong  \bigoplus_{\lambda' \in Z\cdot
\lambda} \cH_{\lambda'}
 \otimes \CC^{m_{\lambda'}},
 \na
 where $ \cH_{\lambda'}$ is the irreducible positive energy
 modules of $\widehat{L\tilde G}$ at level $k$ with highest weight
 $ \lambda' $ and $m_{\lambda'}=1$ except for
 $\widehat{L(PSO(4n))}_{\chi, -}$ (Cf. Remark \ref{PSO})
  with  $Z\cdot\lambda = \{\lambda\}$, $k$ is even
  in which case $m_{\lambda} =2$.  Moreover, the group of
  discontinuous loops corresponding to elements in $Z_\lambda$
  acts on $\cH^\rho_{Z\cdot \lambda}$ via the character $\rho$.
  Note that,  if $Z_\lambda$ is non-trivial, 
   $\cH^\rho_{Z\cdot \lambda}$ ($\rho \in Hom (Z_\lambda,
  U(1))$) have the same Virasoro   character
  \[
  \sum_{\lambda'\in Z\cdot \lambda} \chi_{k, \lambda'}(\tau).
  \]
The appearance of the character $\rho \in Hom (Z_\lambda,
  U(1))$ in the representation of $\widehat{LG}_\chi$ should be
  understood in terms of the  Borel-Weil theory for loop groups (Cf. \cite{PS}).

 Denote by $R_{k, \chi}(LG)$ the Abelian group
  generated by the  irreducible positive energy
 representations of $ \widehat{LG}_\chi$.
 Denote by $\chi_{k, Z \lambda}^\rho $
 the Kac-Peterson character corresponding to the irreducible positive energy
 representation $\cH_{Z\cdot \lambda}^\rho$ of $\widehat{LG}_\chi$ for $Z\cdot \lambda 
 \in \Lambda_{k, \chi}^*$, and $\rho \in Hom (Z_\lambda, U(1))$. 
 Then $R_{k, \chi}(LG)$
 is an Abelian group generated by those $\chi_{k, Z \lambda}^\rho $.

\subsection{Multiplicative bundle gerbes}

Multiplicative bundle gerbes on $G$ have
  transgressive   Dixmier-Douady class (\cite{CJMSW}).
For a compact, connected and simple Lie group $G$,
$H^4(BG, \ZZ)\cong \ZZ$, in terms of the generators of
$H^4(BG, \ZZ)$ and $H^4(B\tilde G, \ZZ)$, a level $k \in \ZZ \cong H^4(B\tilde G, \ZZ)$
is transgressive for $G= \tilde{G}/Z$, if and only if  (Cf. \cite{DijWit} and \cite{MooSei})
\ba\label{quantization:condition}
\disp{\frac k2}<\lambda_{i(z)}, \lambda_{i(z)}> \in \ZZ,
\na
where $\{\lambda_{i(z)}\}_{z \in Z}$ are those special fundamental co-weights
corresponding to  elements in $Z$. Let $\ell_m$ be the smallest positive integer
such that all transgressive levels  for $G= \tilde{G}/Z$ is a multiple of $\ell_m$.
We call $\ell_m$ the multiplicative level of $G$.

We let $\ell_f$ be the smallest positive integer
lying in the image of $H^3(G, \ZZ) \to H^3(\tilde G, \ZZ)$,
it is called the fundamental level of $G$ in \cite{Tol}, see also \cite{FGK}.  Note that the
fundamental level for $SO(3)$ is $2$, and the multiplicative level of $SO(3)$ is $4$.

We  need to construct a $G$-equivariant   bundle gerbe over $G$ for
a multiplicative level $k\in \ell_m\ZZ$. Note that the multiplicative
level $\ell_m$ is always a multiple of the basic level $\ell_b$.  Given $k \in \ell_m\ZZ$
and $ \chi \in Hom (Z, U(1))$, there exists a canonical central extension
of $LG$ given by (\ref{LG:extension}).

Let $\cP G$  be the space of smooth maps $f: \RR \to G$ such that
    $ \theta \mapsto f(\theta +2\pi) f(\theta)^{-1}$ is constant. The map $\cP G \to G$
    given by $f\mapsto f(2\pi)f(0)^{-1}$ defines a
principal $LG$-bundle over $G$. Then we can follow the construction in the proof of
Proposition \ref{bg:G-equivariant} to define a $G$-equivariant   bundle gerbe over $G$.

 \begin{proposition} Given a level $k \in \ell_b\ZZ$ and $\chi \in Hom (Z, U(1))$,
 the lifting bundle  gerbe associated to the central extension $\widehat{LG}_\chi$
 as in  (\ref{LG:alpha}) and the principal
 $LG$-bundle $\cP G$ over $G$ is a $G$-equivariant bundle gerbe over $G$, denoted by
 $\cG_{(k, \chi), G}$, whose equivariant Dixmier-Douady class
 is determined by $(k, \chi)$.
 \end{proposition}
 \begin{proof} It is easy  to show that
 the lifting bundle  gerbe associated to the central extension $\widehat{LG}_\chi$
 as in  (\ref{LG:alpha}) and the principal
 $LG$-bundle $\cP G$ over $G$ is a $G$-equivariant bundle gerbe. Pull-back
 $\widehat{LG}_\chi$ to  $L_Z\tilde G$, we get a central extension
 of $L_Z\tilde G$ which is classified by the
 level $k$, and a cocycle in $H^2(Z, U(1))$.    From the
 exact sequence,
 \[ 0 \to Ext(G, U(1)) \longrightarrow  H^3_G(G, \ZZ) \longrightarrow  H^3(G, \ZZ),
 \]
 we know that the
  equivariant Dixmier-Douady class of $\cG_{(k, \chi), G}$, as a class in
  $H^3_G(G, \ZZ)$, is canonically determined by $(k, \chi)$.
  \end{proof}

    We keep the notation $\cG_k = \cG_{k, \tilde G}$ for the bundle
   gerbe of level $k$ on the simply connected Lie group $\tilde G$
   whose equivariant Dixmier-Douady class
 is $k$ times the generator  in $H^3_{\tilde G}(\tilde G, \ZZ)$.

  \subsection{$G$-equivariant bundle gerbe modules}

  Let $\Theta_k$ be  the closed bi-invariant 3-form on $G$,
\[
\Theta_k =  \disp{\frac {k}{12} <\theta, [\theta, \theta]>
     =    \frac {k}{12} <\bar \theta, [\bar \theta, \bar\theta]>},
     \]
where $\theta, \bar{\theta} \in \Omega^1(G, \g)$ the left- and right-
   invariant Maurer-Cartan forms.
    Note that  $k$ has to be  a multiple of
    the fundamental level $\ell_f$ in order that $\Theta_k$ defines an integral
    cohomology class in $H^3(G, \ZZ)$. Here we require that $k$ is a multiple
    of the basic level $\ell_b$, note that $\ell_b$ is a multiple of $\ell_f$ (Cf. \cite{Tol}).

   Group-valued moment maps for quasi-Hamiltonian $G$-manifolds
   and the corresponding Hamiltonian $LG$-manifolds at level $k$
have been studied also
   for compact, semi-simple, non-simply connected Lie groups (\cite{AAM}).
A quasi-Hamiltonian
   manifold $(M, \omega, \mu)$ and its Hamiltonian $L G$-manifold
  $(\hat{M}, \hat{\omega}, \hat{\mu})$ at level $k$ give rise to the following diagram
  \ba\label{quasi:non-simply}
  \xymatrix{ \hat{M}\ar[d]_{\pi} \ar[r]^{\hat{\mu}} & L\g^* \ar[d]^{Hol}\\
     M \ar[r]^{\mu} &G,
     }
 \na
 where $\hat{\mu }:   \hat{M} \to L\g^*$ is the moment map for the
   Hamiltonian $LG$-action.  To be precise, $(\hat{M}, \hat{\omega}, \hat{\mu})$
   is actually a Hamiltonian $\widehat{L G}_\chi$-manifold with  $\widehat{L G}_\chi$-equivariant
   moment map
   $$\hat{\mu }:   \hat{M} \longrightarrow  L\g^* = L\g^* \times \{ k \}
   \hookrightarrow L\g^* \oplus \RR,  $$
   and $L\g^*$ is again identified as the $L^2_p$-connections on the principal
   $G$-bundle over $S^1$.

 The proof of the following proposition is straightforward,
see the proof of Theorem
 \ref{bg:module:rank1}.

 \begin{proposition} Given a level $k\in \ell_b\ZZ$ and $\chi \in Hom(Z, U(1))$,
 an $\widehat{LG}_\chi$-equivariant vector bundle $\cE$ over a Hamiltonian
  $LG$-manifold $\hat{M}$ defines  a  $G$-equivariant bundle gerbe module of the
  $G$-equivariant bundle gerbe $\cG_{(k, \chi),  G}$. That means, the corresponding
 quasi-Hamiltonian $G$-manifold is a generalized equivariant bundle gerbe
 $D$-brane of $\cG_{(k, \chi),  G}$.
 \end{proposition}

Let $\cQ_{(k, \chi),  G}$
be the category of $G$-equivariant bundle gerbe modules of
$\cG_{(k, \chi),  G}$. Then the quantization functor defined as in
Definition \ref{quantization} can be carried over to $\cQ_{(k, \chi),  G}$.  The
coadjoint orbits of the affine $LG$-action on $L\g^*$ through
$\lambda \in \Lambda_{k, \chi}^*$ provide  examples in $\cQ_{(k, \chi),  G}$.

  \begin{remark}
 Given  a Riemann surface $\Sigma_{g, 1}$ of genus $g$ with only
one boundary component (which is pointed
 by fixing a base point on the boundary), the moduli space $\hat{\cM}_{\Sigma_{g, 1}}$ of
 flat connections on $\Sigma_{g, 1}\times G$ modulo
 those gauge transformations which  are trivial on the boundary
 is only a Hamiltonian $L_0G$-manifold at level $k$, for a transgressive
 level $k$. The holonomy map 
 $$ \hat{\cM}_{\Sigma_{g, 1}} /\Omega_0 G$$
 defines a quasi-Hamiltonian $G$ manifold. But   $\hat{\cM}_{\Sigma_{g, 1}}$  does {\bf  not}
      admit a $LG$-action.
\end{remark}

 Given a  Hamiltonian $L_0G$-manifold $\hat{M}$ at level $k$ with a free 
 $\Omega_0G$-action such that
 $\hat{M}/\Omega_0G$ is a quasi-Hamiltonian  $G$-manifold,
      we need to construct in a canonical way  a
   Hamiltonian $LG$-manifold $\hat{M}^\#$ at level $k$, and an associated
   principal $\Omega G$-bundle over $M$, i.e. fill in the diagram,
 \[
 \xymatrix{ \hat{M}^\# \ar[d]_{\Omega G}  \ar[r]^{\hat{\mu}^\#}
   & L\g^* \ar[d]_{\Omega G} \\
   M  \ar[r]^{\mu} &  G.
     }
      \]
      
      \begin{lemma} \label{L0G2LG} Given a  Hamiltonian $L_0G$-manifold $\hat{M}$ at level $k$ with
      its quasi-Hamiltonian $G$-manifold $\mu: M \to G$, then
      the fiber product $\tilde{M} = M\times_G\tilde G$ is a quasi-Hamiltonian $\tilde G$-manifold with
      the corresponding Hamiltonian $L\tilde G$-manifold $\hat{M}^\#$ is also a
       Hamiltonian $LG$-manifold.
      \end{lemma}
      \begin{proof} Define $\tilde G$-action on $M\times_G\tilde G$
      via
      \[
      \tilde{g}\cdot (m, \tilde{g_1}) = (\pi(\tilde{g})\cdot m, \tilde{g}  \tilde{g_1}\tilde{g}^{-1}),
      \]
      where $\pi: \tilde{G} \to G$ is the covering map. Then the projection
      $\tilde{\mu}: \tilde{M} =  M\times_G\tilde G\to \tilde{G}$ is a $\tilde G$-equivariant
      map.  $M$ is a quasi-Hamiltonian $G$-manifold, it is easy to see
      that  $(\tilde{M}, \tilde{\mu})$ is a quasi-Hamiltonian $\tilde G$-manifold and the following
      diagram commutes:
      \ba\label{1}\xymatrix{ \tilde{M} \ar[d]  \ar[r]^{\tilde{\mu} }
   & \tilde{G} \ar[d]_{\pi} \\
    M  \ar[r]^{ \mu} &  G.
     }
      \na
        As $\tilde{G}$ is simply-connected, the corresponding Hamiltonian $L\tilde G$-manifold
      is given by the fiber product $\hat{M}^\# = \tilde{M}\times_{\tilde{G}} L\g^*$:
      \ba\label{2}
      \xymatrix{ \hat{M}^\# \ar[d]  \ar[r]^{\hat{\mu}^\#}
   & L\g^* \ar[d]_{Hol} \\
   \tilde{M}  \ar[r]^{\tilde{\mu}} &  \tilde{G},
     }
      \na
      where $Hol:  L\g^* \to \tilde{G}$ is a universal $\Omega\tilde{G}$-bundle over $\tilde{G}$.
      Composing Diagram (\ref{1}) and Diagram (\ref{2}), we see
      that $\hat{M}^\# = \tilde{M}\times_{\tilde{G}} L\g^* $ is a Hamiltonian
      $LG$-manifold with $LG$-action given by the affine coadjoint action
      of $LG$ on $L\g^*$ at level k:
      \[
        \xymatrix{ \hat{M}^\# \ar[d]_{\Omega G}  \ar[r]^{\hat{\mu}^\#}
   & L\g^* \ar[d]_{\Omega G} \\
   M  \ar[r]^{\mu} &  G.
     }
      \]
      \end{proof}

 \begin{remark}\label{subtle} \begin{enumerate}
  \item Note that the center $Z\subset \tilde{G}$
  acts trivially on $\tilde{M}$, and $Z\subset L\tilde {G}$, as constant
  loops,  acts trivially on $\hat{M}^\#$. 
    Using the surjection $L\tilde G \to L_0G$, a Hamiltonian
 $LG$-manifold $\hat{M}^\#$ at level $k$ admits
 a Hamiltonian  $L\tilde G$-action at level $k$.
\item Suppose  $\hat{M}^\#$ is  pre-quantizable as a Hamiltonian $LG$-manifold
with a $\widehat{LG}_\chi$-equivariant pre-quantization line $\cL_{\hat{M}^\#}$, then
$\hat{M}^\#$ is also  pre-quantizable  as a Hamiltonian $L\tilde{G}$-manifold
and  the $\widehat{L\tilde{G}}$-equivariant  line bundle $\cL_{\hat{M}^\#}$
on which the center $Z\subset L\tilde {G}$ acts via the
character $\chi$.
 This defines a natural map
 \ba
 \pi_{k, \chi}^D: \cQ_{(k, \chi), G} \longrightarrow
 \cQ_{k, \tilde G}.
 \na
 \end{enumerate}
\end{remark}

  Notice that the coadjoint orbit of the affine coadjoint $LG$-action on $L\g^*$ consists
  of $\{\zeta_z|z\in Z\}$-orbit,  where the distinguished discontinuous loops
  $\{\zeta_z\}$ become   smooth loops in $LG$,  the affine coadjoint action of these smooth loops
      on $L\g^*$ is exactly the  $Z$-action defined
  by (\ref{action:affine}). 
We fix a representative $\lambda$ in each $Z$-orbit $Z \cdot \lambda
\subset \Lambda^*_k$.

  Given a  Hamiltonian $LG$-manifold $(\hat{M}^\#, \hat{\mu}^\#) $ at level $k$ with
a $\widehat{LG}_\chi$-equivariant pre-quantization line $\cL_{\hat{M}^\#}$, 
from Lemma \ref{L0G2LG} and Remark \ref{subtle}, we know that a  Hamiltonian $LG$-manifold 
  $(\hat{M}^\#, \hat{\mu}^\#) $ at level $k$ is also a Hamiltonian $L\tilde{G}$-manifold at level
  $k$ and its
  pre-quantization line bundle  $\cL_{\hat{M}^\#}$ is $\widehat{L\tilde{G}}$-equivariant. The
  Hamiltonian $L\tilde{G}$-reduction at $\lambda$
\ba\label{reduction:diff}
 \hat{M}^\#_{\lambda, \tilde G}:
  = (\hat{\mu}^\#)^{-1}(L\tilde{G}\cdot  \lambda)/ L{\tilde G}  
  \cong (\hat{\mu}^\#)^{-1}(\lambda)/(L{\tilde G})_\lambda ,
\na 
with its  pre-quantization line bundle is given by
\[
  \cL_{\lambda, \tilde G}:= \cL_{\hat{M}^\#}|_{(\hat{\mu}^\#)^{-1} (\lambda) }
  \times_{(\widehat{L\tilde{G}})_{ \lambda }}
 \CC_{(*\lambda,   1)},
\] 
where the action of 
$\widehat{L\tilde{G}}_{\lambda}$  on $\CC_{(*\lambda,   1)}\cong \CC$ 
is given by the weight $(\lambda, 1)$, notice
that for $\lambda \in \Lambda_{k, \chi}^*$,
 the weight $(+\lambda, 1)$  agrees with the character $\chi$ when
restricted to $Z\subset  \widehat{L\tilde{G}}_{\lambda}$.

The  Hamiltonian $LG$-reduction at $\lambda \in \Lambda_{k, \chi}^*$,
\[
\hat{M}^\#_{\lambda, G}:= (\hat{\mu}^\#)^{-1}(LG\cdot \lambda )/LG
 \cong (\hat{\mu}^\#)^{-1}(\lambda )/(LG)_\lambda,
\]
doesn't depend on the choice of $\lambda$ in its $Z$-orbit. Here 
$(LG)_\lambda$ denotes the isotropic group of $LG$-action at $\lambda$. The pre-quantization line
bundle  over $\hat{M}^\#_\lambda$ depends on a choice of a
character $\rho \in Hom (Z_\lambda, U(1))$ if 
\[
Z_\lambda =\{ z \in Z| z\cdot \lambda =\lambda\}
\]
is non-trivial. For each character $\rho \in Hom (Z_\lambda, U(1))$, the
corresponding pre-quantization line
bundle  over $\hat{M}^\#_{\lambda, G}$ is given by
\[
\cL_{\lambda, G}^\rho := 
\cL_{\hat{M}^\#}|_{(\hat{\mu}^\#)^{-1} (\lambda) }\times_{(\widehat{LG})_{ \lambda }}
 \CC_{(*\lambda, \rho^{-1}, 1)},
\] 
where $*\lambda$ is the dominant weight of the irreducible representation of $G$
  dual to the one with weight $\lambda$,  the action of 
$\widehat{LG}_{\lambda}$  on $\CC_{(*\lambda, \rho^{-1}, 1)}\cong \CC$ is determined by
the action of $(\widehat{L_0G})_\lambda$ on $\CC$ via the weight $(*\lambda, 1)$
and the character $\rho^{-1}$ through the exact sequence
\[
1\to (\widehat{L_0G})_\lambda\longrightarrow 
(\widehat{LG})_{ \lambda } \longrightarrow Z_\lambda \to 1,
\]
where the quotient group $\widehat{LG}_{ \lambda }/\widehat{L_0G}_\lambda$
 is identified with the group of distinguished
 loops defined by (\ref{distinguished})  corresponding to elements in the 
stabilizer $Z_\lambda$.   We denote  by $\dirac_{\lambda, G}^\rho$ 
the $Spin^c$ Dirac operator 
associated to the line bundle $\cL_\lambda^\rho$. It is understood that
when $Z_\lambda$ is trivial, then
$$ \cL_{\lambda, G}^\rho = \cL_{\lambda, G}^{1} \quad and \quad
 \dirac_{\lambda, G}^\rho=  \dirac_{\lambda, G}^{1}. $$ 

The following proposition identifies the spaces of sections of the line
bundles  $\cL^\rho_{\lambda, G}$ and  $\cL_{\lambda, \tilde G}$, denoted by
$\Gamma (\cL^\rho_{\lambda, G})$ and $\Gamma (\cL_{\lambda, \tilde G})$ respectively.

\begin{proposition} \label{section:lambda} Given $\lambda\in \Lambda_{k, \chi}^*$, the space of
sections of the line bundle $\cL^\rho_{\lambda, G}$ consists of sections for
the line bundle $\cL_{\hat{M}^\#}|_{(\hat{\mu}^\#)^{-1} (\lambda) }$ over
$(\hat{\mu}^\#)^{-1} (\lambda)$ with weight $(\lambda, \rho, 1)$ for
 the action of $(\widehat{LG})_{ \lambda }$; and the 
space of sections of the line bundle $\cL_{\lambda, \tilde G}$ consists of sections of
the line bundle $\cL_{\hat{M}^\#}|_{(\hat{\mu}^\#)^{-1} (\lambda) }$ over
$(\hat{\mu}^\#)^{-1} (\lambda)$ with weight $(\lambda,  1)$  for the action of
$(\widehat{L\tilde{G}})_{ \lambda }$. Moverover, 
\[
\Gamma (\cL_{\lambda, \tilde G}) \cong \bigoplus_{\rho\in Hom (Z_\lambda, U(1))}
\Gamma (\cL^\rho_{\lambda, G}).
\]
\end{proposition}
\begin{proof} We know that  $\hat{M}^\#_{\lambda, \tilde G}$
has $|Z_\lambda|$-components,  on which $Z_\lambda$ acts transitively via the group
of distinguished discontinuous loops.  Each component of $\hat{M}^\#_{\lambda, \tilde G}$
is diffeomorphic to $\hat{M}^\#_{\lambda, G}$.

  The line bundle $\cL_{\hat{M}^\#}|_{(\hat{\mu}^\#)^{-1} (\lambda) }$
is $(\widehat{LG})_{ \lambda }$-equivariant and $(\widehat{L\tilde{G}})_{ \lambda }$-equivariant.
From the definition of $\cL_{\lambda, \tilde G}$, we can see
that   the  space of sections of the line bundle $\cL_{\lambda, \tilde G}$ consists of sections of
the line bundle $\cL_{\hat{M}^\#}|_{(\hat{\mu}^\#)^{-1} (\lambda) }$ over
$(\hat{\mu}^\#)^{-1} (\lambda)$ with weight $(\lambda,  1)$  for the action of
$(\widehat{L\tilde{G}})_{\lambda }$;  and similarly the space of
sections of the line bundle $\cL^\rho_{\lambda, G}$ consists of sections for
the line bundle $\cL_{\hat{M}^\#}|_{(\hat{\mu}^\#)^{-1} (\lambda) }$ over
$(\hat{\mu}^\#)^{-1} (\lambda)$ with weight $(\lambda, \rho, 1)$ for
 the action of $(\widehat{LG})_{ \lambda }$.

There is a $Z_\lambda$-covering map
\[
\pi:\hat{M}^\#_{\lambda, \tilde G}   \longrightarrow  \hat{M}^\#_{\lambda, G},
\]
 and there is a bundle isomorphism between  $\cL_{\lambda, \tilde G}$ and $\pi^*\cL^\rho_{\lambda, G}$.
For a section $s\in  \Gamma (\cL^\rho_{\lambda, G})$, the linear map
 $ s \mapsto  \pi^*s$
 identifies $\Gamma (\cL^\rho_{\lambda, G})$ with a subspace of $\Gamma (\cL_{\lambda, \tilde G})$
 such that 
 \[
 \Gamma (\cL_{\lambda, \tilde G}) \cong \bigoplus_{\rho\in Hom (Z_\lambda, U(1))}
\Gamma (\cL^\rho_{\lambda, G}).
\]
\end{proof}

  \begin{definition} \label{quantization:non} Given a $G$-equivariant bundle gerbe module
     $(\hat{M}^\#,  \cE) \in  \cQ_{(k, \chi),  G}$, we define
the  quantization of $(\hat{M}^\#,  \cE)$  to be
to be
\[
 \chi_{(k, \chi), G} (\hat{M},  \cE)
 = \sum_{Z \cdot \lambda\in \Lambda_{k, \chi}^*/Z}
 \sum_{\rho\in Hom  (Z_\lambda, U(1))} 
Index (\dirac^\rho_{\lambda, G}\otimes \cE, \hat{M}^\#_{\lambda, G})
  \chi_{k, Z\cdot \lambda}^\rho \in R_{k, \chi}(LG).
  \]
\end{definition}

This gives rise to a quantization functor  $  \chi_{(k, \chi), G} :
\cQ_{(k, \chi),  G} \to R_{k, \chi}(LG)$.

\subsection{The fusion category of bundle gerbe modules}

Now recall that our fusion product in the simply connected case uses a
pre-quantizable line bundle over the moduli space $\hat{\cM}_{\Sigma_{0, 3}}$.
For a non-simply connected Lie group, the moduli space $\hat{\cM}_{\Sigma_{0, 3}}$
is quantizable for any transgressive level $k$.

\begin{proposition} The $G$-equivariant bundle gerbe $\cG_{(k,  \chi), G}$
over $G$ is a $G$-equivariant multiplicative bundle gerbe if
$k$ is transgressive and
$\chi$ is the trivial homomorphism.
\end{proposition}
\begin{proof} The first statement
holds from the main result of \cite{CJMSW}.
To be $G$-equivariant and multiplicative,
the central extension
 $\widehat{LG}_\chi$ of $LG$
has to be $G$-equivariant as a principal $U(1)$-bundle
over $LG = G \times \Omega G$
with $G$-action on $\Omega G$ given by conjugation, and under
the face operators from $\pi_i: G\times G \to G$ where
$\pi_0(g_1, g_2) = g_2$, $\pi_1(g_1, g_2) = g_1g_2$ and $\pi_2(g_1, g_2) = g_1$ for
  $(g_1, g_2) \in G \times G$,  there is a $G$-equivariant stable isomorphism
  \[
  \pi_0^*\cG_{ (k,  \chi), G} \otimes \pi_2^*\cG_{ (k,  \chi), G}
  \longrightarrow \pi_1^*\cG_{ (k,  \chi), G}.
  \]
These conditions hold if  $\chi \in Hom (Z, U(1))$ is the trivial homomorphism.
\end{proof}

This proposition determines the
conditions under which
we may obtain the fusion category of bundle
gerbe modules in this non-simply connected situation.
From now on in this subsection, we assume that the level $k$ is
multiplicative for $G$, i.e., $k\in \ell_m\ZZ$, this excludes the case of
$\cQ_{(k, \chi, -), G}$ for $G = PSO(4n)$, as $\chi$ has to be non-trivial
for $\widehat{LG}_{\chi, -}$.

We can define the fusion of two quasi-Hamiltonian $G$-manifolds
   $\mu_i: ( M_i, \omega_i ) \to G$ ($i=1, 2$) as
   \ba\label{fusion:G}
   (M_1, \omega_1, \mu_1)\boxtimes (M_2, \omega_2, \mu_2) =
(M_1 \times M_2,   \mu_1\cdot \mu_2),
\na
with the 2-form $\omega_1 + \omega_2 + \disp{\frac 12} <\mu^*_1\theta, \mu_2^*\bar{\theta}>$.
Then $(M_1 \times M_2, \mu_1\cdot \mu_2) $
is again a  quasi-Hamiltonian $G$-manifold, and the corresponding Hamiltonian
 $LG$-manifold at level $k$ is given by
 \ba\label{fusion:LG}
 \widehat{M_1\boxtimes  M_2} = (M_1 \times M_2)\times_G L\g^*,
 \na
Where the universal principal $\Omega G$-bundle over $G$ factors through $\tilde G$: $L\g^*\to
\tilde{G} \to G$.

For the Hamiltonian $L_0G$-manifold  $\hat{\cM}_{\Sigma_{g, 1}}$ at any
multiplicative level $k$,
the proof of Proposition \ref{moduli:gb-module}
can be adapted to show that
$\hat{ \cM}^\#_{\Sigma_{g, 1}}$ admits a $\widehat{LG}_1$
pre-quantization line bundle. The key point in the proof is the
Segal-Witten reciprocity property for transgressive central extensions
(we omit the details).

The category of $G$-equivariant bundle gerbe modules for  $\cG_{(k, 1), G}$,
when $k $ is transgressive for $G$,
admits  the fusion object  given by the fiber product
\[
\hat{\cM}^\#_{\Sigma_{0,3}} = \hat{\cM}_{\Sigma_{0,3}} \times_{G^3}\tilde G^3,
\]
which is a Hamiltonian $(LG)^3$-manifold at level $k$ with a $\widehat{(LG)}_1^3$-equivariant
pre-quantization line bundle $\cL_{\Sigma_{0,3}}$.  This can be verified
 the fact that  $\widehat{M_1\boxtimes M_2}$ defined
by (\ref{fusion:LG}) is diffeomorphic to the Hamiltonian quotient
\[
\bigl((\hat{M}_1 \times \hat{M_2} )\times \hat{\cM}^\#_{\Sigma_{0,3}}\bigr)//
diag(LG\times LG).
\]
Moreover, if  $\hat{M_1}$ and $\hat{M_2}$ admit  $\widehat{ LG }_1$-equivariant
pre-quantization line bundles $\cL_1$ and $\cL_2$ respectively, then
\[
\bigl((\cL_1 \times  \cL_2  )\times  \cL_{\Sigma_{0,3}}\bigr)//
\widehat{diag (LG)^2}_1
\]
defines a $\widehat{LG}_1$-equivariant pre-quantization line
bundle over $\widehat{M_1\boxtimes M_2} $. Hence, the fusion
category of rank one $G$-equivariant bundle gerbe modules of
$\cG_{(k, 1), G}$ is well defined. We  denote this fusion
category by $(\cQ_{ (k, 1), G}, \boxtimes_G)$.

We can apply the quantization functor $\chi_{(k, 1), G}: (\cQ_{
(k, 1), G}, \boxtimes_G) \longrightarrow  R_{ k,1} (LG)$ to define
a fusion product on  $R_{ k,1} (LG)$. Note that irreducible
positive energy representations  in $R_{ k,1} (LG)$ are labelled
by 
\[
\{(Z\cdot\lambda, \rho_\lambda) | Z\cdot\lambda \subset \Lambda_{k, 1}^*,
\rho_\lambda \in Hom (Z_\lambda, U(1))\}.
\]

\begin{definition}
We define the fusion coefficient for $R_{ k,1} (LG)$ as:
\ba\label{fusionring:G}
N_{(Z\cdot\lambda, \rho_\lambda), (Z\cdot\mu, \rho_\mu)}^{(Z\cdot\nu, \rho_\nu)}
:= Index \bigl(\dirac_{(*\lambda, *\mu, \nu), G}^{(\rho_\lambda, \rho_\mu, \rho_\nu)},
\hat{\cM}^\#_{\Sigma_{0, 3}} (G, *\lambda, *\mu, \nu)\bigr),
\na
where $\hat{\cM}^\#_{\Sigma_{0, 3}} (G, *\lambda, *\mu, \nu)$ denotes the Hamiltonian
$(LG)^3$-reduction of $\hat{\cM}^\#_{\Sigma_{0, 3}}$ at $(*\lambda, *\mu, \nu)$, and 
$\dirac_{(*\lambda, *\mu, \nu), G}^{(\rho_\lambda, \rho_\mu, \rho_\nu)}$ is the corresponding
$Spin^c$ Dirac operator.
\end{definition}

\begin{theorem}\label{main:non}
$\chi_{k, Z\cdot\lambda}^{\rho_\lambda} \ast  \chi_{k, Z\cdot\mu}^{\rho_\mu} 
=\sum_{(Z\cdot\nu, \rho_\nu)} 
N_{(Z\cdot\lambda, \rho_\lambda), (Z\cdot\mu, \rho_\mu)}^{(Z\cdot\nu, \rho_\nu)} 
 \chi_{k, Z\cdot\nu}^{\rho_\nu} $ defines a fusion ring structure on $R_{ k,1} (LG)$ with the
 unit given by $\chi_{k, Z\cdot 0}$, the representation corresponding to the $Z$-orbit through $0$. 
\end{theorem}
\begin{proof}
The fusion product defined on the $\cQ_{(k, 1), G}$ is commutative and associative modulo
$LG$-equivariant symplecmorphisms and equivalence of $\widehat{LG}_1$-equivariant line bundles. This
imply that the fusion product on  $R_{ k,1} (LG)$ is commutative and associative. The unit 
in $\cQ_{(k, 1), G}$ is given by $\Omega G$ with its pre-quantization line
bundle $\widehat{\Omega G}_1 \times_{U(1)}\CC$, the quantization functor 
\[
\chi_{(k, 1), G}:  \cQ_{(k, 1), G}  \longrightarrow  R_{ k,1} (LG)
\]
sends $\Omega G$ to $\chi_{k, Z\cdot 0}$.
\end{proof}

\begin{proposition}
If $Z\cdot \lambda$,
$Z\cdot \mu$ and $Z\cdot \nu$ are free $Z$-orbits, then 
\[
N_{ Z\cdot\lambda,  Z\cdot\mu }^{ Z\cdot\nu } = \sum_{z\in Z} N_{\lambda, \mu}^{z\cdot\nu}.
\]
\end{proposition}
\begin{proof}  Note that  the Verlinde coefficients $\{N_{\lambda, \mu}^\nu\}$ for $R_k(L\tilde G)$
  satisfy the following symmetry under the action
of $Z$:
\[
N_{z_1\cdot\lambda, z_2\cdot \mu}^{z_1z_2 \cdot
\nu} = N_{\lambda,  \mu}^{\nu}
\]
for any $z_1, z_2  \in Z$. This is due to the fact that
the moduli spaces for calculating the Verlinde coefficients
$N_{\lambda,  \mu}^{\nu}$ and $N_{z_1\cdot\lambda,
z_2\cdot \mu}^{z_1z_2 \cdot \nu}$  for $R_k(L\tilde{G})$ are identical:
\[
\hat{\cM}^\#_{\Sigma_{0,3}}(\tilde{G},*\lambda, *\mu, \nu) \cong
\hat{\cM}^\#_{\Sigma_{0,3}}(\tilde{G},*(z_1 \cdot\lambda),
*(z_2\cdot\mu),z_1z_2\cdot \nu)
\]
from the holonomy descriptions of these moduli spaces.  These facts imply that
\[
\hat{\cM}^\#_{\Sigma_{0, 3}} (G, *\lambda, *\mu, \nu) \cong 
\bigsqcup_{z\in Z} \hat{\cM}^\#_{\Sigma_{0, 3}} (\tilde{G}, *\lambda, *\mu, z\cdot \nu),
\]
as symplectic manifolds, and their corresponding pre-quantization line bundles are also
equivalent for free $Z$-actions on  $Z\cdot \lambda$,
$Z\cdot \mu$ and $Z\cdot \nu$. Hence, we have 
\[
N_{ Z\cdot\lambda,  Z\cdot\mu }^{ Z\cdot\nu } = \sum_{z\in Z} N_{\lambda, \mu}^{z\cdot\nu}.
\]
\end{proof}

\begin{remark} The fusion category $(\cQ_{(k, 1), G}, \boxtimes_G)$ is actually a braided tensor
category, see \cite{BK} for a definition of  a braided tensor
category, where the braiding isomorphism  for two Hamiltonian $LG$-manifolds $\hat{M_1}$ and
$\hat{M_2}$ 
\[ 
\hat{M_1}\boxtimes \hat{M_2} \longrightarrow \hat{M_2}\boxtimes \hat{M_1} 
\]
is induced by a diffeomorphism of $\Sigma_{0, 3}$
exchanging the two incoming boundaries.    Applying  the conformal
model for $\Sigma_{0,3}$
\[
P_{w, q , q } = \{ z\in \CC|  |q | \leq |z| \leq 1, |z-w| \geq |q |\}
\]
with boundary points $1$, $q $ and $w+q $, where $0< |w| < 1$, and $0<|q | < |w|-|q | < 1-2
|q |$. Then the conformal
model for $\Sigma_{0,3}$ with two incoming boundaries exchanged is given by
$P_{-w, q, q}$. 
Note that $P_{w, q , q }$ and $P_{-w, q, q}$ are connected by the path 
$P_{e^{i\theta}w, q , q }$ for $\theta \in [0, \pi]$.
The Pentagon axiom, Triangle axiom and Hexagon axioms
follow from the multiplicative property of the equivariant bundle
gerbe $\cG_{(k, 1), G}$.
This braiding isomorphism
is important to determine the fusion coefficients for $R_{ k,1} (LG)$
involving $Z$-orbits with non-trivial stabilizer.
\end{remark}

To illustrate our result, we end this section by a detailed
study for $G=SO(3)$ and $G= SU(3)/\ZZ_3$. 

\subsection{An example for $G=SO(3)$}

 Note that $SO(3)= SU(2)/\ZZ_2$, the basic level is $2$, and the level is
transgressive if and only if it is a multiple of $4$. 
Given a class $(k, \chi) \in 2\ZZ \oplus (\ZZ_2, U(1))$ where $\chi = \pm 1 \in \ZZ_2$, we have
the corresponding equivariant bundle gerbes
$\cG_{k, \pm 1}$ for $k = 4n$ or $k= 4n+2$ ($n>0$). We first give a complete classification of
all irreducible positive energy representations of $\widehat{LSO(3)}_{\chi}$ at level $k\in 2\ZZ$
and $\chi = \pm 1$.

For $k = 4n$ and $\chi = +1$, the irreducible positive energy representations
of  $\widehat{LSO(3)}_{+1}$ are labelled by
$\ZZ_2 = Z(SU(2))$-orbits in the space of level $k$ dominant weights. We instead use the
half-integers (half weights) $j = 0, 1/2, 1, 3/2, \cdots, 2n-1/2, 2n$ to
label level $k$ dominant
weights of $LSU(2)$. Denote
\[
\cH_{0}, \cH_{1/2}, \cH_{1}, \cdots, \cH_{2n-1/2}, \cH_{2n}
\]
the corresponding   irreducible positive energy representations of $\widehat{LSU(2)}$ at level $4n$.
These representations of $\widehat{LSU(2)}$ can be obtained by
(geometric) quantization of equivariant bundle gerbe $D$-branes given by
conjugacy classes labelled by those half-integer  representations of $SU(2)$, or
equivalently, quantization of  equivariant bundle gerbe modules of the corresponding
affine coadjoint $LSU(2)$-orbits at level $4n$.

Then the irreducible positive energy representations
of  $\widehat{LSO(3)}_{+1}$, as $ \widehat{LSU(2)}$-modules,  are given by
\[\begin{array}{c}
\cH_{0}\oplus \cH_{2n}, \\[2mm] \cH_{1}\oplus \cH_{2n-1}, \\[2mm]
\cH_{2}\oplus \cH_{2n-2}, \\[2mm]\qquad  \vdots \\  \cH_{n-1}\oplus \cH_{ n+1},\\[2mm]
 \cH_{n }^{\pm},
\end{array}
\]
where the spin $n$   is the fixed point of the $\ZZ_2$-action: $j
\mapsto 2n-j$, and $\cH^\pm_n \cong \cH_n$ as a
$\widehat{LSU(2)}$-module, with the group of discontinuous loops
corresponding to $\ZZ_2$ acting via the character $\pm 1$. These
representations can be thought of as quantization of the
projection of $\ZZ_2$-orbits of those conjugacy classes in $SU(2)$
with integer spin weights.

It is straightforward to verify that $R_{4n, +1}(LSO(3))$ admits a fusion product,
from which we obtain the non-diagonal modular invariant
(diagonal modular invariant for an extension of the chiral algebra by $Z$
\[
\cZ_{4n, +1} = \sum_{j= 0}^{k/4 -1 } |\chi_{k, j} + \chi_{k, k/2 -j}|^2 + 2|\chi_{k, n} |^2.
\]
This agrees with the formula from fixed point resolution for simple current extensions
in \cite{SY} \cite{CIZ}.

For $k = 4n$ and $\chi =-1$, the irreducible positive energy representations
of  $\widehat{LSO(3)}_{-1}$, as $ \widehat{LSU(2)}$-modules,  are given by
\[\begin{array}{l}
\cH_{1/2}\oplus \cH_{2n-1/2},\\[2mm] \cH_{3/2}\oplus \cH_{2n-3/2}, \\[2mm]
\cH_{5/2}\oplus \cH_{2n-5/2},\\[2mm] \qquad \vdots \\ \cH_{n-1/2}\oplus \cH_{ n+1/2}.
\end{array} \]

Similarly, for $k = 4n+2$ and $\chi = +1$,
the irreducible positive energy representations
of  $\widehat{LSO(3)}_{+1}$, as $ \widehat{LSU(2)}$-modules,  are given by
\[\begin{array}{c}
\cH_{0}\oplus \cH_{2n+1}, \\[2mm] \cH_{1}\oplus \cH_{2n }, \\[2mm]
\cH_{2}\oplus \cH_{2n-1},\\[2mm]\qquad  \vdots \\  \cH_{n-1}\oplus \cH_{ n+2},\\[2mm] \cH_{n }\oplus \cH_{ n+1 }.
\end{array} \]
For $k = 4n+2$ and $\chi =-1$, the irreducible positive energy representations
of  $\widehat{LSO(3)}_{-1}$, as $ \widehat{LSU(2)}$-modules,  are given by
\[\begin{array}{c}
\cH_{1/2}\oplus \cH_{2n+1/2},\\[2mm] \cH_{3/2}\oplus \cH_{2n-1/2},\\[2mm]
\cH_{5/2}\oplus \cH_{2n-3/2},\\[2mm]\qquad  \vdots \\ \cH_{n-1/2}\oplus \cH_{ n+3/2},\\[2mm]
 \cH_{n+1/2}^\pm,
\end{array} \]
where the spin $n+1/2$ is the fixed point of the $\ZZ_2$-action:
$j \mapsto 2n+1 -j$, $\cH^\pm_{n+ 1/2} \cong \cH_{n+ 1/2}$ as a
$\widehat{LSU(2)}$-module, with the group of discontinuous loops
corresponding to $\ZZ_2$ acting via the character $\pm 1$.

For each $k \in 2\ZZ$ and $\chi \in \ZZ_2 \cong Hom (\ZZ_2, U(1))$, the quantization
functor
\[
\chi_{(k, \chi), SO(3)}: \cQ_{(k, \chi), SO(3)} \longrightarrow R_{k, \chi} (LSO(3)),
\]
from the  category  $\cQ_{(k, \chi), SO(3)}$ of $SO(3)$-equivariant  generalized bundle gerbe
modules of $\cG_{k, \chi}$ over $SO(3)$ to $R_{k, \chi} (LSO(3))$, is surjective. Among the
four cases discussed above, $\cQ_{(k, \chi), SO(3)}$ admits a fusion product
structure if $k = 4n$ and $\chi =+1$, and the corresponding quantization
functor $$\chi_{(4n, +1), SO(3)}:\cQ_{(4n , +1), SO(3)} \to R_{4n, +1}(LSO(3))$$
 preserves the fusion products.

\subsection{An example for $G=SU(3)/\ZZ_3$}

 Denote by
\[
\{\a_1, \a_2\}, \quad and \quad \{ \a_1^\vee, \a_2^\vee\}
\]
the simple roots and the simple co-roots of $SU(3)$ respectively.
The fundamental weights and co-weights are denoted by
\[
\{\lambda_1, \lambda_2\}, \quad and \quad \{\lambda^\vee_1,
\lambda_2^\vee \}
\]
respectively. Then we know that the highest root is given by $\a_1
+\a_2$ and
\[
\left\{ \begin{array}{l} \a_1^\vee + \a_2^\vee = \lambda_1^\vee
+\lambda_2^\vee,\\[2mm]
\a_1^\vee - \a_2^\vee = 3(\lambda_1^\vee -\lambda_2^\vee),
\end{array}
\right.
\]
which gives the isomorphism $$\Lambda_w^\vee /\Lambda_r^\vee \cong
\ZZ (\lambda_1^\vee -\lambda_2^\vee)/3\ZZ(\lambda_1^\vee
-\lambda_2^\vee)  \cong \ZZ_3.$$

The set of  dominant weights at level  $\Lambda_k^*$ is given by
\[
\{ k_1 \lambda_1 + k_2\lambda_2 | k_i \geq 0, k_1 + k_2 \leq  k\},
\]
with the action of $\ZZ_3$ generated by
\[
 k_1 \lambda_1 + k_2\lambda_2 \mapsto (k-k_1 - k_2)\lambda_1 +
 k_1 \lambda_2.
 \]
 It is easy to see that the character map
 $ \Lambda_k^* \to Hom(\ZZ_3, U(1))$ is induced
 by $$k_1 \lambda_1 + k_2\lambda_2 \mapsto (k_1 -k_2) (mod \  3),$$
 and $\ZZ_3$-action admits a fixed point if and only if
 $k \in 3\ZZ$ with the fixed point given by $\frac k3 \lambda_1 +
 \frac k3 \lambda_2$.

 For $G=SU(3)/\ZZ_3$,  as the multiplicative level for $SU(3)/\ZZ_3$ is $3$, we know that
the transgressive level for $SU(3)/\ZZ_3$ is given by $k\in 3\ZZ$, we know that the multiplicative
  bundle gerbes  over $G$ are classified by their
  Dixmier-Douady classes: the level $k \in 3\ZZ$. We note that $\ell_b =1$ for $SU(3)/\ZZ_3$.
  Hence, the equivariant bundle gerbe $\cG_{(k, \chi),SU(3)/\ZZ_3}$  exists for any
  level $k\in \ZZ$ and $\chi \in Hom(\ZZ_3, U(1))$. Here we only consider
  the transgressive levels.

  Given $k\in 3\ZZ$ and $\chi \in \ZZ_3 \cong Hom (\ZZ_3, U(1))$, we have the
corresponding quantization functor:
\[
\chi_{(k/3, \chi), SU(3)/\ZZ_3}: \cQ_{(k/3, \chi), SU(3)/\ZZ_3}
\longrightarrow R_{k, \chi} (L(SU(3)/\ZZ_3)).
\]
Denote by $\cH_{(k_1, k_2)}$ the positive energy irreducible
representation of $\widehat{LSU(3)}$ of the highest weight $k_1
\lambda_1 + k_2\lambda_2 \in \Lambda_k^*$. Then
 $R_{k, \chi} (L(SU(3)/\ZZ_3))$ is generated by
 \[
 \cH_{(k_1, k_2)} \oplus \cH_{(k-k_1-k_2, k_1)} \oplus \cH_{(k_2,
 k-k_1-k_2)},
 \]
 for $k_1\lambda_1 +k_2 \lambda_2 \in \Lambda_k^*$ and
$\chi ([\lambda_1^\vee - \lambda_2^\vee ]) = e^{2\pi i \frac { k_1 - k_2 }{3}}$ for
$k_1\lambda_1 +k_2 \lambda_2$ with trivial stabilizer. For $\chi=1$ and $k_1 =k_2 =k/3$,
there are three additional representations:
\[
\cH_{(k/3, k/3)}^{\rho_0}, \quad \cH_{(k/3, k/3)}^{\rho_1}, \quad  \cH_{(k/3, k/3)}^{\rho_1},
\]
which are equivalent to $\cH_{(k/3, k/3)}$ as $ \widehat{LSU(3)}$-modules, with 
with the group of discontinuous loops
corresponding to $\ZZ_3$ acting via the character $\rho_i \in Hom (\ZZ_3, U(1))$.

We give a complete list for $k=3$ and $k=6$ as follows:

\begin{enumerate}
\item For $k=3$, if $\chi ([\lambda_1^\vee - \lambda_2^\vee ]) =1$, then
$R_{3, \chi} (L(SU(3)/\ZZ_3))$ is generated by
 \[\begin{array}{c}
 \cH_{(0, 0)} \oplus \cH_{(3, 0)} \oplus \cH_{(0, 3)},\\[2mm]
 \cH_{(1, 1)}^{\rho_0}, \quad  \cH_{(1, 1)}^{\rho_1}, \quad  \cH_{(1, 1)}^{\rho_2}, 
 \end{array}
 \]
 as  $ \widehat{LSU(3)}$-modules, where $ (1, 1)$ is the fixed point of the $\ZZ_3$-action,
  $\cH^{\rho_i}_{(1, 1)} \cong \cH_{(1,1)}$ as a
$\widehat{LSU(3)}$-module, with the group of discontinuous loops
corresponding to $\ZZ_2$ acting via the character $\rho_i$;
 if $\chi ([\lambda_1^\vee - \lambda_2^\vee ]) =e^{2\pi i/3}$, then
 $R_{3, \chi} (L(SU(3)/\ZZ_3))$ is generated by
 \[\cH_{(1, 0)} \oplus \cH_{(2, 1)} \oplus \cH_{(0, 2)},
 \]
 as  $ \widehat{LSU(3)}$-modules;
 if $\chi ([\lambda_1^\vee - \lambda_2^\vee ]) =e^{4\pi i/3}$, then
 $R_{3, \chi} (L(SU(3)/\ZZ_3))$ is generated by
 \[\cH_{(2, 0)} \oplus \cH_{(1, 2)} \oplus \cH_{(0,  1)},
 \]as  $ \widehat{LSU(3)}$-modules. Note that the Verlinde ring
 $R_{3, +1} (L(SU(3)/\ZZ_3))$ gives rise to a
 modular invariant for $SU(3)/\ZZ_3$ at level $3$ (Cf. \cite{Ber}):
 \[
 \cZ_{(3, +1), SU(3)/\ZZ_3}
 = |\chi_{3, (0, 0)}  +  \chi_{3, (3, 0)}  +  \chi_{3, (0,3)}|^2
 + 3|\chi_{3, (1, 1)}|^2.
 \]
 \item For $k=6$,
 if $\chi ([\lambda_1^\vee - \lambda_2^\vee ]) =1$, then
$R_{6, \chi} (L(SU(3)/\ZZ_3))$ is generated by
 \[\begin{array}{c}
 \cH_{(0, 0)} \oplus \cH_{(6, 0)} \oplus \cH_{(0, 6)},\\[2mm]
 \cH_{(1, 1)} \oplus\cH_{(4, 1)} \oplus\cH_{(1, 4)},\\[2mm]
 \cH_{(2, 2)}^{\rho_0}, \quad \cH_{(2, 2)}^{\rho_1}, \quad
 \cH_{(2, 2)}^{\rho_2},  \\[2mm]
 \cH_{(3, 3)} \oplus\cH_{(0, 3)} \oplus\cH_{(3, 0)},
 \end{array}
 \]
 as  $ \widehat{LSU(3)}$-modules, where $ (2, 2)$ is the fixed point of the $\ZZ_3$-action,
  $\cH^{\rho_i}_{(2, 2)} \cong \cH_{(2,2)}$ as a
$\widehat{LSU(3)}$-module, with the group of discontinuous loops
corresponding to $\ZZ_2$ acting via the character $\rho_i$;
 if $\chi ([\lambda_1^\vee - \lambda_2^\vee ]) =e^{2\pi i/3} $, then
$R_{6, \chi} (L(SU(3)/\ZZ_3))$ is generated by
 \[\begin{array}{l}
 \cH_{(1, 0)} \oplus \cH_{(5, 1)} \oplus \cH_{(0, 5)},\\[2mm]
 \cH_{(2, 1)} \oplus\cH_{(3, 2)} \oplus\cH_{(1, 3)},\\[2mm]
 \cH_{(4, 0)} \oplus\cH_{(2, 4)} \oplus\cH_{(0, 2)},
 \end{array}
 \]
 as  $ \widehat{LSU(3)}$-modules;
if $\chi ([\lambda_1^\vee - \lambda_2^\vee ]) = e^{4\pi i/3}$, then
$R_{6, \chi} (L(SU(3)/\ZZ_3))$ is generated by
 \[\begin{array}{l}
 \cH_{(2, 0)} \oplus \cH_{(4, 2)} \oplus \cH_{(0, 4)},\\[2mm]
 \cH_{(3, 1)} \oplus\cH_{(2, 4)} \oplus\cH_{(1, 2)},\\[2mm]
 \cH_{(5, 0)} \oplus\cH_{(1,5)} \oplus\cH_{(0, 1)},
 \end{array}
 \]
 as  $ \widehat{LSU(3)}$-modules

 Note that   only
 $R_{6, +1} (L(SU(3)/\ZZ_3))$  admits a ring structure, which
  gives rise to a   modular invariant for $SU(3)/\ZZ_3$ at level $6$ (Cf. \cite{BG})
 \[\begin{array}{lll}
&&  \cZ_{(3, +1), SU(3)/\ZZ_3} \\[2mm]
 &=&  |\chi_{6, (0, 0)}  +  \chi_{6, (6, 0)}  +  \chi_{6, (0,6)}|^2
+  |\chi_{6, (1, 1)}  +  \chi_{6, (4, 1)}  +  \chi_{6, (1,4)}|^2 \\[2mm]
&&+|\chi_{6, (3, 3)}  +  \chi_{6, (0, 3)}  +  \chi_{6, (3,0)}|^2
 + 3|\chi_{6, (2, 2)}|^2.
 \end{array}
 \]
 \end{enumerate}


\begin{thebibliography}{9999}

  \bibitem{AAM} A.  Alekseev, A.  Malkin and E. Meinrenken {\sl
   Lie group valued moment maps.}
 J. Differential Geom. 48 (1998), no. 3, 445--495.

 \bibitem{AleSch}  A.Yu. Alekseev, V. Schomerus {\sl D-branes in the WZW model},
 Phys.Rev. D60 (1999) 061901.

 \bibitem{Ati} M. Atiyah {\sl The geometry and physics of knots}, Cambridge University
 Press, 1990.

  \bibitem{AB}  M. Atiyah,   R. Bott {\sl  The Yang-Mills equations over
Riemann surfaces.} Philos. Trans. Roy. Soc. London Ser. A 308 (1983), no. 1505, 523--615.

\bibitem{AS}  M. Atiyah, G. Segal {\sl Twisted $K$-theory}, preprint.

\bibitem{BG} E. Baver, D. Gepner {\sl  Fusion rules for extended current algebras}.  
  Modern Phys. Lett. A 11 (1996), no. 24, 1929--1945.

\bibitem{BK} B. Bakalov, A. Kirillov, Jr. {\sl Lectures on tensor categories and modular functors},
University Lecture Series, Vol. 21.


 \bibitem{Bea} A. Beauville  {\sl Conformal blocks, fusion rules and the Verlinde formula},
  (Ramat Gan, 1993),   No. 9,  Israel Math. Conf. Proc.,  1996, 75--96.

 \bibitem{Ber} D. Bernard, {\sl String characters from Kac-Moody automorphisms},
  Nucl. Phys. B288,    1987,  628-648, 

\bibitem{BCMMS} P. Bouwknegt, A. Carey, V. Mathai, M.  Murray, D.
 Stevenson {\em Twisted $K$-theory and
 $K$-theory of bundle gerbes.}  Comm. Math. Phys.  228  (2002),  no. 1, 17--45.

\bibitem{BDR}  P. Bouwknegt, P. Dawson, D. Ridout  {\sl
D-branes on group manifolds and fusion rings, } JHEP 0212 (2002) 065.

  \bibitem{BryMcL} J-L. Brylinski, D. A. McLaughlin
  {\sl The converse of the Segal-Witten reciprocity law},
  Internat. Math. Res. Notices  1996,  no. 8, 371--380.

 \bibitem{CIZ} A. Cappelli, C. Itzykson, J. B. Zuber {\sl The A-D-E classification of minimal
 and $A^{(1)}_1$ conformal invariant theories}.  Comm. Math. Phys., 113, 1-26 (1987)

\bibitem{Car} J. Cardy {\sl  Boundary conditions, fusion rules and the Verlinde formula},
Nuclear Phys. B 324 (1989), no. 3, 581--596.

\bibitem{CJMSW}  A. L. Carey, S. Johnson, M. K. Murray, D. Stevenson,
  B. L. Wang {\sl Bundle gerbes for Chern-Simons and Wess-Zumino-Witten
  theories}, to apeear in Comm. Math. Phys., math.DG/0410013.

\bibitem{CMM} A.L. Carey,  M.K. Murray and J. Mickelsson
{\sl Bundle gerbes in
quantum field theory}  Rev Math Phys
 {\bf 12} (2000) 65-90.

\bibitem{CW1} A.L. Carey, B.L. Wang {\sl The Universal Gerbe and Local Family Index Theory},
 preprint, math.DG/0407243

\bibitem{CW2} A.L. Carey, B.L. Wang {\sl Multiplicative bundle
gerbe and equivariant twisted K-theory}, preprint.


  \bibitem{DijWit}
    Dijkgraaf, R and Witten, E.,
  {\sl Topological Gauge
  Theories and Group Cohomology}
  Commun. Math. Phys. {\bf 129},
  393--429, (1990).


\bibitem{Don} S. Donaldson {\sl Boundary value problems for Yang-Mills fields.}
   Jour. of Geom. Phys. 8 (1992), 89-122.

\bibitem{EFK} P. Etingof, I. Frenkel, A. Kirillov, {\sl
Spherical functions on affine Lie groups.}
Duke Math. J. 80 (1995), no. 1, 59--90.

\bibitem{FFFS}
G. Felder, J. Frohlich,  J. Fuchs, C.  Schweigert {\sl
The geometry of WZW branes.} J. Geom. Phys. 34 (2000), no. 2, 162--190.

\bibitem{FuSS} J. Fuchs, U. Ray, B. Schellekens, C. Schweigert,
{\sl Twining characters and orbit Lie algebras}, hep-th/9612060.


 \bibitem{FGK} G. Felder, K. Gawedzki, A. Kupiainen
 {\sl Spectra of Wess-Zumino-Witten models with arbitrary simple groups.}
 Comm. Math. Phys. 117 (1988), no. 1, 127--158.

\bibitem{FHT} D. Freed, M.J. Hopkins and C. Teleman {\sl
Twisted K-theory and Loop Group Representations, }  math.AT/0312155,  preprint.

\bibitem{FHT2} D. Freed, M.J. Hopkins and C. Teleman {\sl
Twisted equivariant K-theory with complex coefficients,} math.AT/0206257, preprint.

\bibitem{FS} S. Fredenhagen, V. Schomerus {\sl
Branes on Group Manifolds, Gluon Condensates, and twisted K-theory.}
JHEP 0104 (2001) 007.

 \bibitem{FreWit} D. Freed and E. Witten {\sl  Anomalies in string theory with $D$-branes.}
    Asian J. Math. 3 (1999), no. 4, 819--851.


\bibitem{GGR}  M. Gaberdiel, T. Gannon, D. Roggenkamp {\sl
The D-branes of SU(n),} JHEP 0407 (2004) 015.

  \bibitem{Gaw0} K. Gawedzki{\sl Topological actions in two-dimensional quantum field theory
 }, in  Nonperturbative Quantum Field Theories, ed. G. 't Hooft,
  A. Jaffe, G. Mack, P. K. Mitter, R. Stora, NATO Series vol. 185, Plenum Press (1988), 101--142.


    \bibitem{Gaw} K. Gawedzki {\sl Abelian and non-Abelian branes in WZW models and gerbes},
    hep-th/0406072.

 \bibitem{GawRei}  K. Gawedzki and N. Reis {\sl WZW branes and gerbes.}
    Rev. Math. Phys.  14  (2002),  no. 12, 1281--1334.

\bibitem{GW} R. Goodman, N. Wallach, {\sl
 Structure and unitary cocycle representations of loop groups and the group of
  diffeomorphisms of the circle.} J. Reine Angew. Math. 347 (1984), 69--133.


 \bibitem{LauXu} C. Laurent-Gengoux and P. Xu {\sl  Quantization of
 pre-quasi-symplectic groupoids and their Hamiltonian spaces, }    423--454, Progr. Math., 232,  2005.
 
 \bibitem{LW}  W. Lerche, J. Walcher {\sl Boundary Rings and N=2 Coset Models}, 
 Nucl.Phys. B625 (2002) 97-127. hep-th/0011107.



\bibitem{MMS} J. Maldacena,  N.  Seiberg, G.  Moore {\sl 
Geometrical interpretation of D-branes in gauged WZW models. }
J. High Energy Phys. 2001, no. 7. hep-th/0105038.

 \bibitem{MatSte} V. Mathai, D. Stevenson
 {\sl Chern character in twisted K-theory: equivariant and holomorphic cases},
 Commun.Math.Phys. 228 (2002) 17-49.

\bibitem{Mei} E. Meinrenken {\sl The basic gerbe over a compact simple Lie group}, math.DG/0209194.


  \bibitem{MeiWoo} E. Meinrenken and C. Woodward {\sl Hamiltonian loop group
   actions and Verlinde factorization.}
    J. Differential Geom.  50  (1998),  no. 3, 417--469.

\bibitem{MeiWoo1} E. Meinrenken and C. Woodward
{\sl Fusion of Hamiltonian loop group manifolds and cobordism,} dg-ga/9707019

\bibitem{MooSei}
G. Moore and N. Seiberg, {\sl Taming the conformal Zoo}, Phys. Lett. B  220  (1989),
no. 3, 422--430.

\bibitem{Mur}
M. K. Murray  {\sl Bundle gerbes,}
J. London Math. Soc. (2) {\bf 54}
(1996), no.~2, 403--416.


 \bibitem{PS} A. Pressley and G. Segal {\sl Loop groups},
    Oxford University Press, Oxford,  1988.


  \bibitem{RSW} T. R. Ramadas, I. M.  Singer and J. Weitsman  {\sl
Some comments on Chern-Simons gauge theory.}
Comm. Math. Phys. 126 (1989), no. 2, 409--420.


\bibitem{Sch1} S.  Schafer-Nameki {\sl 
 K-theoretical boundary rings in N = 2 coset models.}
Nucl.\ Phys.\ B {\bf 706} (2005) 531. 
 hep-th/0408060.

\bibitem{Sch2} S.  Schafer-Nameki {\sl D-branes in N = 2 coset models and twisted equivariant K-theory}.
 hep-th/0308058.

 \bibitem{SY} A. Schellekens, S.  Yankielowicz  {\sl Simple currents, modular invariants and
 fixed points.} Internat. J. Modern Phys. A 5 (1990), no. 15, 2903--2952.



\bibitem{Seg} G. Segal  {\sl The definition of conformal field theory,}
  421--577, London Math. Soc. Lecture Note Ser., 308, Cambridge Univ. Press, Cambridge, 2004.



\bibitem{Tel} C. Teleman  {\sl The quantization conjectures revisited, }
Ann, of Math. (2), 152 (1), 1-43, 2000.

\bibitem{Tol} V. Toledano Laredo {\sl
Positive energy representations of the
loop groups of non-simply connected Lie groups.}
Comm. Math. Phys. 207 (1999), no. 2, 307--339.
math.QA/0106196



 \bibitem{Ver}  E. Verlinde  {\sl  Fusion rules and modular
 transformations in $2d$  conformal field theory}, Nuclear Phys. B 300 (1988) 360--376.

 \bibitem{Witt} E. Witten {\sl Non-abelian bosonization in two dimensions},
  Comm. Maths. Phys. 2 (1984), 455-472.

  \bibitem{Wit} E. Witten {\sl Two dimensional gauge theories revisited}, Jour. of
  Geom. Phys.  9 (1992), 303-368.

\bibitem{Wit1} E. Witten {\sl  $D$-branes and $K$-theory. }
J. High Energy Phys. 1998, no. 12, Paper 019.

\bibitem{Wood} N. Woodhouse, {\sl Geometric quantization}, Oxford University Press, 1980.

\bibitem{Xu} P. Xu  {\sl Moment maps and Morita equivalence}, preprint, math.SG/0307319.




  \end{thebibliography}
  \end{document}